\documentclass[twocolumn,trackchanges,twocolappendix]{aastex63}

\hypersetup{linkcolor=red,citecolor=blue,filecolor=cyan,urlcolor=blue}%draft,
\received{May 27, 2021}
\revised{\today}
\accepted{\today}
\submitjournal{ApJS}

\shorttitle{TASOC Light Curve Correction Pipeline}
\shortauthors{Lund et al.}
\newcommand{\ie}{i.e.\@\xspace} %id est
\newcommand{\eg}{e.g.\@\xspace} %id est
\usepackage{xspace}
\newcommand{\kp}{\emph{Kepler}\xspace}
\newcommand{\SC}{$120$ s\xspace}
\newcommand{\SCC}{$20$ s\xspace}
\newcommand{\LC}{$1800$ s\xspace}
\newcommand{\LCC}{$600$ s\xspace}
\newcommand{\code}[1]{\texttt{#1}}
\newcommand{\numax}{\ensuremath{\nu_{\rm max}}\xspace}
\newcommand{\dnu}{\ensuremath{\Delta\nu}\xspace}

\newcommand{\eqref}[1]{Equation~\ref{#1}}
\newcommand{\fref}[1]{Figure~\ref{#1}}

\newcommand{\sref}[1]{Section~\ref{#1}}

\usepackage{lipsum}  
\usepackage{soul}
\usepackage{float}

\usepackage{lineno}
\makeatletter
\newcommand\footnoteref[1]{\protected@xdef\@thefnmark{\ref{#1}}\@footnotemark}
\makeatother

\defcitealias{PaperIII}{Paper III}
\defcitealias{PaperI}{Paper I}
\begin{document}

\title{TESS Data for Asteroseismology: Light Curve Systematics Correction}

\correspondingauthor{Mikkel N. Lund}
\email{mikkelnl@phys.au.dk}

\author[0000-0002-0786-7307]{Mikkel~N.~Lund}
\affil{Stellar Astrophysics Centre, Department of Physics and Astronomy, Aarhus University, Ny Munkegade 120, DK-8000 Aarhus C, Denmark}

\author[0000-0002-0786-7307]{Rasmus~Handberg}
\affil{Stellar Astrophysics Centre, Department of Physics and Astronomy, Aarhus University, Ny Munkegade 120, DK-8000 Aarhus C, Denmark}

\author[0000-0002-1988-143X]{Derek~L.~Buzasi}
\affil{Department of Chemistry and Physics, Florida Gulf Coast University, 10501 FGCU Blvd. S., Fort Myers, FL 33965 USA}

\author[0000-0003-1001-5137]{Lindsey~Carboneau}
\affil{Department of Chemistry and Physics, Florida Gulf Coast University, 10501 FGCU Blvd. S., Fort Myers, FL 33965 USA}
\affil{School of Physics and Astronomy, University of Birmingham, Edgbaston, Birmingham B15 2TT, UK}
\affil{Stellar Astrophysics Centre, Department of Physics and
11
 Astronomy, Aarhus University, Ny Munkegade 120, DK-8000 Aarhus C, Denmark}

\author[0000-0002-0468-4775]{Oliver~J.~Hall}
\affil{European Space Agency (ESA), European Space Research and Technology Centre (ESTEC), Keplerlaan 1, 2201 AZ Noordwijk, The Netherlands}
\affil{School of Physics and Astronomy, University of Birmingham, Edgbaston, Birmingham B15 2TT, UK}
\affil{Stellar Astrophysics Centre, Department of Physics and Astronomy, Aarhus University, Ny Munkegade 120, DK-8000 Aarhus C, Denmark}

\author[0000-0002-2157-7146]{Filipe~Pereira}
\affil{Instituto de Astrof\'{\i}sica e Ci\^{e}ncias do Espa\c{c}o, Universidade do Porto, CAUP, Rua das Estrelas, 4150-762 Porto, Portugal}
\affil{Departamento de F\'{\i}sica e Astronomia, Faculdade de Ci\^{e}ncias da Universidade do Porto, Rua do Campo Alegre, s/n, PT4169-007 Porto, Portugal}

\author[0000-0001-8832-4488]{Daniel~Huber}
\affiliation{Institute for Astronomy, University of Hawai`i, 2680 Woodlawn Drive, Honolulu, HI 96822, USA}

\author[0000-0003-3244-5357]{Daniel~Hey}
\affil{School of Physics, Sydney Institute for Astronomy (SIfA), The University of Sydney, NSW 2006, Australia}

\author[0000-0003-2771-1745]{Timothy~Van~Reeth}
\affil{Institute of Astronomy, KU Leuven, Celestijnenlaan 200D, BUS-3001, Leuven, Belgium}

\author{and the T'DA collaboration}
\begin{abstract}
Data from the Transiting Exoplanet Survey Satellite (TESS) has produced of order one million light curves at cadences of \SC and especially \LC for every ${\sim}27$-day observing sector during its two-year nominal mission.
These data constitute a treasure trove for the study of stellar variability and exoplanets. However, to fully utilize the data in such studies a proper removal of systematic noise sources must be performed before any analysis.
The \emph{TESS Data for Asteroseismology} (T'DA) group is tasked with providing analysis-ready data for the TESS Asteroseismic Science Consortium, which covers the full spectrum of stellar variability types, including stellar oscillations and pulsations, spanning a wide range of variability timescales and amplitudes.
We present here the two current implementations for co-trending of raw photometric light curves from TESS, which cover different regimes of variability to serve the entire seismic community.
We find performance in terms of commonly used noise statistics to meet expectations and to be applicable to a wide range of different intrinsic variability types. Further, we find that the correction of light curves from a full sector of data can be completed well within a few days, meaning that when running in steady-state our routines are able to process one sector before data from the next arrives. Our pipeline is open-source and all processed data will be made available on TASOC and MAST.

\end{abstract}

\keywords{techniques: photometric, methods: data analysis, stars: variables: general}

\section{Introduction}\label{sec:intro}

The space-based photometric missions CoRoT \citep{corot}, \kp \citep{Gilliland2010}, and K2 \citep{Howell2014} have over the last decade and a half revolutionized the field of asteroseismology \citep[\eg,][]{chaplin2013,Hekker2017,Bowman2017,Garcia2019}.
The success of these missions can be attributed to their ability to deliver high-quality photometric observations at a high cadence and spanning long continuous baselines. The Transiting Exoplanet Survey Satellite \citep[TESS;][]{Ricker2014} is the most recent mission to satisfy these criteria, and promises to continue the advancement of asteroseismology \citep{Campante2016,Schofield2019}.
As with \kp, the TESS asteroseismic investigation is organized within the TESS Asteroseismic Science Consortium \citep[TASC; see][]{Lund2017}, and the data for asteroseismic analysis is hosted at the TESS Asteroseismic Science Operations Center (TASOC\footnote{\label{note_tasoc}\url{https://tasoc.dk/}}) and the Mikulski Archive for Space Telescopes (MAST\footnote{\label{note_mast}\url{https://archive.stsci.edu/hlsp/tasoc}}: \dataset[https://doi.org/10.17909/t9-4smn-dx89]{https://doi.org/10.17909/t9-4smn-dx89}).

The primary science goal of TESS is to find exoplanets smaller than Neptune that can be characterized in detail from follow-up observations. The main data product obtained to meet the science goal is observations of a pre-selected list of targets at a \SC cadence. In addition to being suited for exoplanet studies, these data are also required for the asteroseismic study of main-sequence solar-like oscillators. 
Of particular importance for the asteroseismic study of evolved stars, such as red giants for galactic archaeology studies \citep{Miglio2013,TESSgiant}, or classical pulsators \citep[][]{Antoci2019,Holdsworth2021}, are the targets observed in the Full-Frame Images \citep[FFIs;][]{SPOC}, which were obtained every \LC during the nominal mission. 

A key distinction between the processing of data from \kp versus that for TESS is that TESS will neither reduce the FFIs nor deliver light curves for all targets. Targets observed at a \SC cadence will be processed to produce light curves, while FFIs will be made available to the community in a calibrated form and light curves delivered for a subset of targets \citep{2020RNAAS...4..201C}. In the extended TESS mission the FFI cadence is reduced to \LCC, and an additional cadence of \SCC has been introduced. 
Hence, to make full use of all the data from TESS there is a need for a flexible, extensible pipeline that can produce analysis-ready data for a variety of astrophysics applications, including exoplanet science, stellar activity, and of course asteroseismology. Such a pipeline, ``the TASOC pipeline'', is provided and run by the coordinated activity ``TESS Data for Asteroseismology'' (T'DA) within TASC (see \sref{sec:tasoc}). In this paper we describe components of the TASOC pipeline dedicated to removing instrumental effects from the raw light curves extracted from data delivered by TESS. Other pipelines for TESS FFI data reduction exist \citep[see, \eg,][]{eleanor,Oelkers2018,qlp2}, but these are mostly focused on exoplanet searches, while the TASOC pipeline seeks to preserve intrinsic stellar variability covering a range of timescales and amplitudes. 

The paper is structured as follows. In \sref{sec:tasoc} we give an overall introduction to the full TASOC pipeline, and then focus on the details of the light curve correction component in \sref{sec:pipe}. \sref{sec:per} details the performance of the light curve correction pipeline, while \sref{sec:format} describes the format of the data products. We conclude and provide an outlook and discuss potential improvements to the pipeline in \sref{sec:out}.

%%%%%%%%%%%%%%%%%%%%%%%%%%%%%%%%%%%%%%%%%%%%%%%%%%%%%%%%%%%%%%%%%
%%%%%%%%%%%%%%%%%%%%%%%%%%%%%%%%%%%%%%%%%%%%%%%%%%%%%%%%%%%%%%%%%
%%%%%%%%%%%%%%%%%%%%%%%%%%%%%%%%%%%%%%%%%%%%%%%%%%%%%%%%%%%%%%%%%
\section{The TASOC Pipeline}\label{sec:tasoc}
The TASOC pipeline is developed by the T'DA group to serve the asteroseismic community of TASC. Specifically, T'DA is responsible for delivering (1) raw photometric time series from FFIs (\LC/\LCC cadence) and Target Pixel Files (TPFs; \SC/\SCC cadence); (2) light curves corrected for systematic signals, ready for asteroseismic analysis; (3) a preliminary classification of the variability/oscillation/pulsation type of each light curve from (1)-(2), enabling each of the working groups (WGs) of TASC, which are organized according to stellar type, to better target their analysis. 

The TASOC pipeline is open-source and available on GitHub\footnote{\url{https://github.com/tasoc/photometry}, \url{https://github.com/tasoc/corrections}} \citep{rasmus_handberg_2021_5153073,rasmus_handberg_2021_5154027}. All data products from the pipeline are available via the TASOC\footnoteref{note_tasoc} database and on MAST\footnoteref{note_mast} (\dataset[https://doi.org/10.17909/t9-4smn-dx89]{https://doi.org/10.17909/t9-4smn-dx89}), see \sref{sec:format} for more details.

\begin{figure*}
    \centering
    \includegraphics[width=0.7\textwidth]{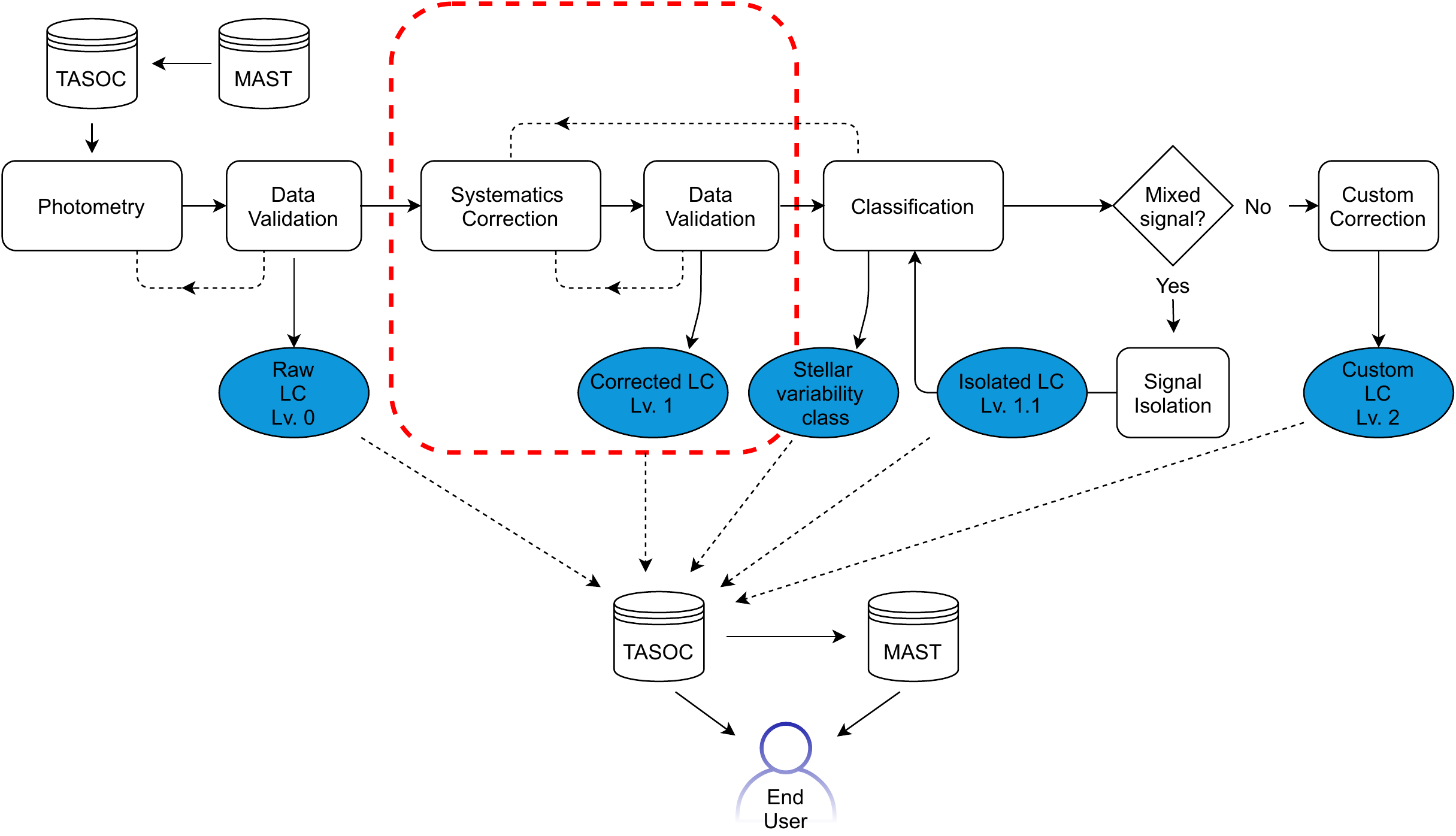}
    \caption{The overall structure of the full TASOC pipeline, with modules given as rectangular boxes, data products as ellipses, and ``TASOC'' and ``MAST'' indicate the databases hosting the data products.   Dashed lines between modules indicate that an iteration might take place. The part enclosed by the red dashed line indicates the pipeline component described in this paper. The ``photometry'' part of the pipeline is described in \citetalias{PaperI}, while the ``classification'' is detailed in \citetalias{PaperIII}.}
    \label{fig:pipe}
\end{figure*}

The photometry module of the pipeline is described in \citet{PaperI} (hereafter \citetalias{PaperI}), and the stellar classification is described in \citet{PaperIII} (hereafter \citetalias{PaperIII}). The current paper deals with the systematics correction of raw light curves. \fref{fig:pipe} gives a general overview of the TASOC pipeline, and the different data products produced. The components enclosed by the red dashed line are those described in this paper.

We note that while the TASOC pipeline is primarily developed with a focus on asteroseismology, the data will also be useful for a wider range of science applications, including exoplanet science, eclipsing binaries, and the study of activity (rotation, flares, etc.).

%%%%%%%%%%%%%%%%%%%%%%%%%%%%%%%%%%%%%%%%%%%%%%%%%%%%%%%%%%%%%%%%%
%%%%%%%%%%%%%%%%%%%%%%%%%%%%%%%%%%%%%%%%%%%%%%%%%%%%%%%%%%%%%%%%%
%%%%%%%%%%%%%%%%%%%%%%%%%%%%%%%%%%%%%%%%%%%%%%%%%%%%%%%%%%%%%%%%%
\section{Light Curve Correction}\label{sec:pipe}

Starting from the raw light curves extracted as detailed in \citetalias{PaperI}, the light curve corrections described in this paper are of a co-trending nature, \ie, using information shared amongst many stars to estimate and account for any systematic signals. For specific types of stars additional detrending may be required to isolate specific signals --- an example could be the filtering of transits for an oscillating exoplanet host, or high-pass filtering of low-frequency variability from stellar rotation.
Depending on the identified variability type, based on the stellar classification module, the TASOC pipeline will at a later stage include a ``custom correction'' making use of classification results to improve the performance of the detrending algorithms and thus produce improved light curves for the different TASC WGs. The modular nature of the pipeline is intended to simplify incorporation of such future improvements and extensions.

Because T'DA serves the entire community within TASC, the light curve correction needs to be able to deal with stars with variability ranging from stochastic solar-like oscillations with a lifetime of minutes, to more coherent pulsators such as $\delta$-Sct/$\gamma$-Dor type, RR-Lyrae stars, and long-period variables, such as Miras and Cepheids. In addition to a large range in typical timescales, the variability amplitudes also range from a few parts-per-million to several magnitudes. The range in timescales and amplitudes (and phase stability) poses a challenge in terms of fully preserving the astrophysical signal, as the level of overlap with the systematic contribution will vary with stellar type.

In the first releases of the correction module, we employ two independent co-trending methodologies. As the quality of the co-trending from the different methods varies with stellar type, we will release both versions. It will then be up to the user to identify which correction is best suited for their specific science case. In a future
release we are hopeful that an informed decision can be made on which correction to use for a 
given star by iterating with the stellar classification module (\citetalias{PaperIII}). 

The two methods incorporated and used for the current release are the
``CBV'' (\sref{sec:cbv}) and the ``Ensemble'' methods (\sref{sec:ens}).
In \sref{sec:per} we will provide a comparison of the two methods to show
typical use-cases where a given method dominates in quality. 
%%%%%%%%%%%%%%%%%%%%%%%%%%%%%%%%%%%%%%%%%%%%%%%%%%%%%%%%%%%%%%%%%
\subsection{CBV fitting method}\label{sec:cbv}

The first of the correction methods employs so-called \textit{co-trending basis vectors} (CBVs) for the correction \citep{kepdatc8}. The CBVs are a set of orthonormal vectors describing the dominant sources of variability shared amongst many stars. The idea is that such shared variability can likely be attributed to systematic signals induced by external sources, rather than representing variability intrinsic to the individual stars.

Below we go through the steps included in generating the CBVs and using them to co-trend the raw TESS light curves.

\subsubsection{CBV generation}\label{sec:cbv_gen}
Many of the procedures used for generating the CBVs are inspired by the pipeline used to process \kp data. For the sake of completeness we go through the steps here, but refer the reader to \citet{kepdatc8} for more details on the specifics of the \kp pipeline. 

\begin{figure}
    \centering
    \includegraphics[width=\columnwidth]{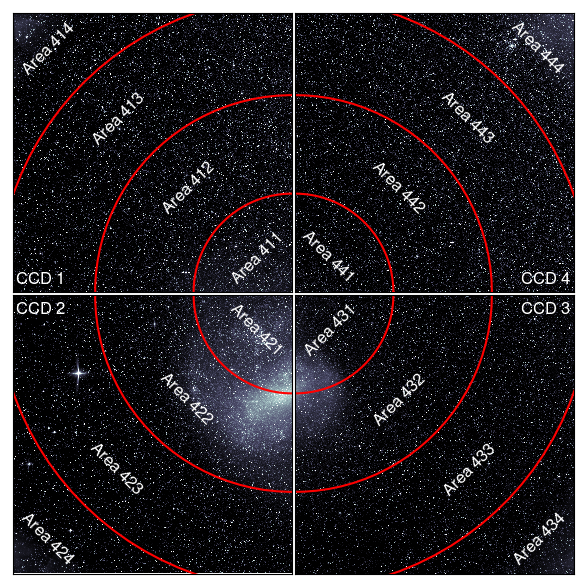}
    \caption{Example of the segmentation of the CCDs into ``CCD Areas''. The example shows the segmentation for camera 4, here during Sector 4. The number of each area is provided in addition to the numbers of the CCDs.}
    \label{fig:ccd}
\end{figure}

The key difference from the approach used for \kp is in the definition of the spatial regions for which the CBVs are computed. In \kp a set of CBVs was computed for each full CCD, and this approach has been carried on to the correction of \SC data from TESS by the Science Processing Operations Center \citep[SPOC;][]{SPOC}. We introduce a subdivision of the TESS CCDs into ``CBV areas'' and compute a set of CBVs for each of these. The areas are currently defined by three concentric circles centered on the camera centers -- in this manner each of the four CCDs per camera will be divided into four areas, giving a total of 16 sets of CBVs per camera. \fref{fig:ccd} gives an illustration of the segmentation for camera 4 (here for observations in Sector 4). The ``Area'' is given by a 3-digit number, where the first gives the camera number, the second gives the CCD number, starting in the top-left and increasing in the anticlockwise direction, and the third gives the region within the CCD, starting from $1$ towards the camera center to $4$ at the camera corner. 
%%%%%%%%%%%%%%%%%%%%%%%%%%%%%%%%%%%%%%%%
%%%%%%%%%%%%%%%%%%%%%%%%%%%%%%%%%%%%%%%%%

The reasoning behind this segmentation is that some systematics, such as focus changes, are expected to depend on the distance to the camera center. By having areas with a difference in their distance to the camera center some of the variation in the systematics should be better captured, allowing for a reduced number of CBVs required in the co-trending, hence reducing the risk of overfitting.

The procedure for generating the CBVs is as follows:
\begin{itemize}
    \item The sample of light curves from a given CCD area is first pruned for targets of high variability, because these can contaminate the identification of shared variability for the CBV generation. A cut is made that deselects targets with a variability larger than $1.3$ times the median variability of all targets in the area. This cut-off value was found to provide a good compromise between deselecting the most highly variable targets, while still retaining a large sample of stars for the CBV generation. We note that the CBV correction will ultimately be applied to all targets. For the measurement of variability we first subtract a 3rd order polynomial fitted using a weighted least squares routine -- the variability is then given by the standard deviation of the residual normalized by the median flux error (as propagated from the photometry pipeline). The 3rd-order polynomial was found to provide a good removal of most long-term variability with a likely systematic origin. Such that, this variability is not used to prune targets for the generation of the CBVs.
    
    \item Data points with certain quality flags or with NaN values in the raw flux time series are removed from the following analysis. Specifically, we remove data points with TESS quality flags 2 (safemode), 8 (Earthpoint), 32 (desaturation event), and 128 (manual exclude), and the potential TASOC quality flag 2 (manual exclude). We note that, at the time of writing, many of the quality flags are not yet populated by the TESS team. Currently we retain data obtained during coarse pointing, as the spike-CBVs (see below) have been found to be able to correct for much of the coarse pointing jitter. We also do not propagate the ``manual exclude'' flags from the \SC data -- in Sector 1 it was found that by propagating these flags about ${\sim}20\%$ of the \LC data of the FFIs would have been discarded.
    
    \item From this reduced ``low variability'' sample we compute the correlation matrix between the median-normalized light curves. Based on the correlation matrix we retain the $50\%$ of targets with the highest median absolute correlation to other targets. The idea behind this selection of targets is that if a light curve is dominated by a systematic variability shared amongst many stars, then their light curves should also show a high level of correlation. For the calculation of the correlation matrix we use Einstein summation in \texttt{Python}, which allows for a very rapid derivation of the correlations even when dealing with several hundred thousand $27$-day light curves. 
    
    \item The light curves from the above step are interpolated using a piecewise cubic hermite interpolating polynomial (PCHIP; as implemented in \texttt{Scipy} \citep{Scipy}) to fill the gaps from masked quality flags or bad datapoints (\eg NaN flux from photometry module) that are individual to a given target -- gaps from shared quality flags are not interpolated. We note that such individual gaps rarely occur, hence generally very few or no points are interpolated. This step ensures that all light curves have the same number of data points, which is required for the following procedures. After the interpolation the light curves are mean normalized. The interpolated points are removed in the final light curves.
    
    \item The Co-trending Basis Vectors (CBVs) are given by the orthonomal basis vectors of a principal component analysis (PCA). The PCA is performed with singular value decomposition (SVD) using the \texttt{Python} package \texttt{scikit-learn} \citep{scikit-learn}. We currently include 16 components in the PCA.

    \item The calculation of the CBVs in the above step is iterated with pruning of single targets with a high weight in a given CBV. From the PCA analysis we can access the weight of each light curve in the construction of a given basis vector. If a given CBV truly represents a shared variability, then a given light curve should not have a weight significantly higher that the other light curves. To identify such potential high-weight targets we compute the entropy of the weights and compare to expectations from an assumed Gaussian entropy distribution with a width similar to standard deviation of the weights \citep[see][for details]{kepdatc8}. A negative difference indicates a poor entropy with dominating contribution for a single or a few stars. Based on testing different limits we find that a value of $-0.5$ provides a good upper limit on the difference -- if this criteria is not met we remove the star with the largest relative weight and re-compute the CBVs. This process is iterated until the calculated entropy meets expectations. Given the pruning of highly variable targets in a previous step, this step typically only removes a few stars, if any at all, from contributing to the CBVs.

    \item The derived CBVs are then tested for having a high signal-to-noise ratio (SNR). Some CBVs, typically the ones with the least information content, will have a large and apparently random scatter. While some of this variability will be truly systematic, the risk of introducing high-frequency noise in the light curve when fitting the CBV outweighs the benefit of including the CBV in the co-trending. We require that in order for a CBV to be considered in the co-trending the power level between the standard deviation of the CBV ($\sigma_{\rm signal}$) and the standard deviation of the first differences of the CBV ($\sigma_{\rm noise}$) should be less than 5 dB, that is, 
    \begin{equation}
        10  \log_{10}( \sigma_{\rm signal}^2 / \sigma_{\rm noise}^2 )\,\,  {\rm dB}< 5
    \end{equation}
    
    \item The high SNR CBVs are then split into components showing low frequency and/or low amplitude variations (we shall refer to these as ``normal'' CBVs) and ``spike'' CBVs that, as the name suggests, contain large single-bin spikes. The spikes are typically remnants of reaction wheel desaturation events (where currently only the central cadence is flagged in \LC data; see \fref{fig:ex_dump}) or times when TESS observed in coarse pointing. The normal and spike CBVs are fitted independently, but always in pairs. To identify and separate out the spikes of a given CBV we first apply a Savitzky-Golay filter to the CBV. This low-pass filtered version of the CBV is then subtracted and the absolute value is taken of the residuals. In these values we then identify peaks that are larger than three times the standard deviation of the absolute residuals (computed using the median absolute deviation). We note that propagating the quality flags from the \SC data (especially the ``manual exclude'' flag) could mitigate a lot of the spikes, though this would be at the cost of discarding many data points.  

\end{itemize}

\begin{figure*}
    \centering
    \includegraphics[width=\textwidth]{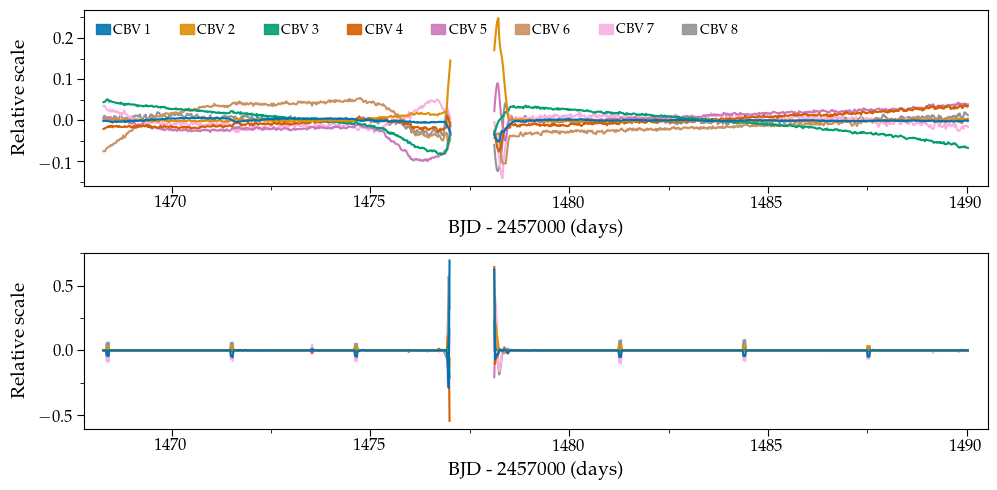}
    \caption{Example of CBVs computed for TESS Sector 6, CCD area $443$. Top: ``Normal'' CBV components. Bottom: ``Spike'' CBV components. The numbers for the different CBVs, increasing with decreasing level of importance, is given with the top panel legend.}
    \label{fig:cbv}
\end{figure*}

\begin{figure*}
    \centering
    \includegraphics[width=\textwidth]{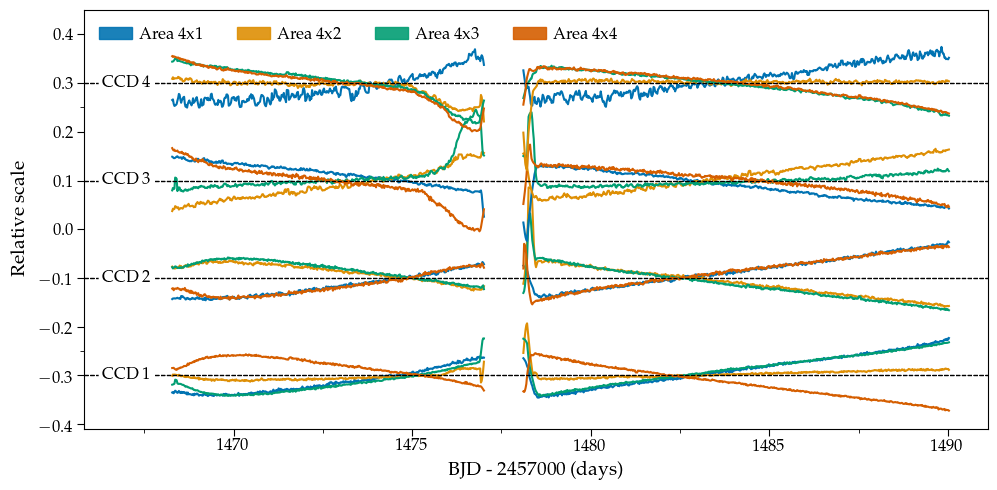}
    \caption{Example of the variation in the third CBV with CCD-area in camera 4, Sector 6. The primary and secondary CBVs mainly focus on the strong systematics on either side of the downlink gap. The different sets of CBVs correspond to the different CCDs, while the color indicates the region within the CCD -- as an example the bottom blue CBV, belonging to CCD 1, comes from CCD Area 411.}
    \label{fig:cbv2}
\end{figure*}

\fref{fig:cbv} gives an example of the computed CBVs from Sector 6 data, showing both the ``normal'' and ``spike'' CBVs. As seen, the normal CBVs typically exhibit a 2-week variation, corresponding to the orbit of TESS. The most dominant contribution to the \LC spike CBVs is the reaction wheel desaturation events every ${\sim}3.125$ days. \fref{fig:cbv2} illustrates the variation in the primary CBV component with position on the camera, showing the importance of segmenting the camera into areas each with their own set of CBVs. Had only a single set of CBVs been calculated for a camera, several components would have been needed to capture the variability of the primary CBV in the segmented setup.
%%%%%%%%%%%%%%%%%%%%%%%%%%%%%%%%%%%%%%%%
%%%%%%%%%%%%%%%%%%%%%%%%%%%%%%%%%%%%%%%%%

%%%%%%%%%%%%%%%%%%%%%%%%%%%%%%%%%%%%%%%%
%%%%%%%%%%%%%%%%%%%%%%%%%%%%%%%%%%%%%%%%%

\subsubsection{CBV fitting}\label{sec:cbv_fit}

The correction of the raw photometry with the CBVs is done by representing the systematic signal to be removed as a linear combination of the CBVs:
\begin{equation}\label{eq:corr}
    {F_{\rm corr}} = {F_{\rm raw}} - \sum_{i=0}^{N} \hat{c}_i \times \textrm{CBV}_i\, ,
\end{equation}
where $i$ runs over the $N$ CBVs included in the correction, and $\hat{c}_i$ is the corresponding scaling coefficient.
The coefficients are given by the ordinary least squares solution as
\begin{equation}\label{eq:fit}
    \hat{\textbf{c}} ={\rm  (\textbf{A}^{T} \textbf{A})^{-1}  \textbf{A}^{T}}\, \textbf{f}\, ,
\end{equation}
where $\textbf{A}$ is the matrix with the normal and spike CBVs as columns and $\textbf{f}$ is the raw flux light curve. In $\textbf{A}$ we also add a column with ones to allow for an offset.

The correction with \eqref{eq:corr} and fitting with \eqref{eq:fit} is done iteratively with a sigma-clipping of the residual corrected flux ($\rm Flux_{corr}$), where points more than $4\sigma$ away are removed (with $\sigma = \rm 1.4826\, MAD\{|Flux_{corr}|\}$). This iteration continues until the fitting coefficients from \eqref{eq:fit} remain unchanged, or a maximum of 50 iterations has been reached -- the maximum number of iterations is rarely reached, but in the odd case it is reached a warning flag is set for the processing of the target. This robust sigma-clipping ensures that outliers that have not been flagged by the quality flags or transient signals (flares, eclipses, transits, etc.) do not influence the fit of the CBVs to the raw light curve. Finally, we use the Bayesian Information Criterion (BIC) to decide on the number $N$ of 16 calculated CBVs to include in the correction.

For the correction of \SC cadence targets we do not compute separate CBVs, but use an interpolated (cubic spline) version of the \LC cadence CBVs. The reason for not generating CBVs based on \SC cadence data is the much reduced number of targets available in a given CBV area, resulting in noisy CBVs that have a tendency to add significant amounts of white noise to the co-trended light curves.

%%%%%%%%%%%%%%%%%%%%%%%%%%%%%%%%%%%%%%%%%%%%%%%%%%%%%%%%%%%%%%%%%
\subsection{Ensemble fitting method}\label{sec:ens}

The second correction method employs ensemble photometry \citep{brown_gilliland1988} to characterize shared variability by constructing an average standard star using a combination of light curves from stars nearby in the field. The approach focuses on characterizing and removing {\em local} systematic effects. Below we describe the steps involved in the Ensemble correction algorithm.

We note that currently the Ensemble method is currently only applied to \LC data because of a lack of sufficient stars for constructing a proper ensemble from \SC data. Using synthetic data, we have conducted numerical experiments with interpolating a \SC ensemble from the \LC data, and results are promising, but not yet ready for inclusion in our final product. Further development of the method to allow the inclusion of \SC stars is underway.

\subsubsection{Ensemble Generation}
\label{sec:ens_gen}
The most important parts of the Ensemble correction method are the determination of appropriate members of the stellar ensemble and the creation of the ``average'' light curve from that group. Here we outline the steps included in that process, which is illustrated in Figure~\ref{fig:ensflow}.
\begin{itemize}
\item Data points in the raw flux time series with NaN values or non-zero quality flags as listed for the CBV method in \sref{sec:cbv_gen} are removed from the succeeding analysis. This typically removes less than 2\% of points, primarily from cadences associated with spacecraft anomalies and scattered light. For the resulting time series we determine three basic statistical parameters for later use in the process. Two of these are the median value and the variability characterized by the range between the 5th and 95th percentile of the flux values ($R_{var}$; see, \eg, \citet{Basri2011}). We also characterize the noise level of the time series using the standard deviation of the {\it differenced} light curve ($\sigma_D$), a technique which accounts for the nonstationary nature of the typical stellar light curve \citep{Nason2006}. We then return a median-divided light curve to use for detrending purposes.

\item Potential ensemble members are identified based on Euclidean distance from the target star. We initially begin with 100 candidates, and further require that they lie on the same camera and CCD as the target. For each potential ensemble member, we calculate the same basic statistical parameters as done for the target, and normalize by the light curve median value.
%%%%%%%%%%%%%%%%%%%%%%%%%%%%%%%%%%%%%%%%
\begin{figure}
    \centering
    \includegraphics[width=0.95\columnwidth]{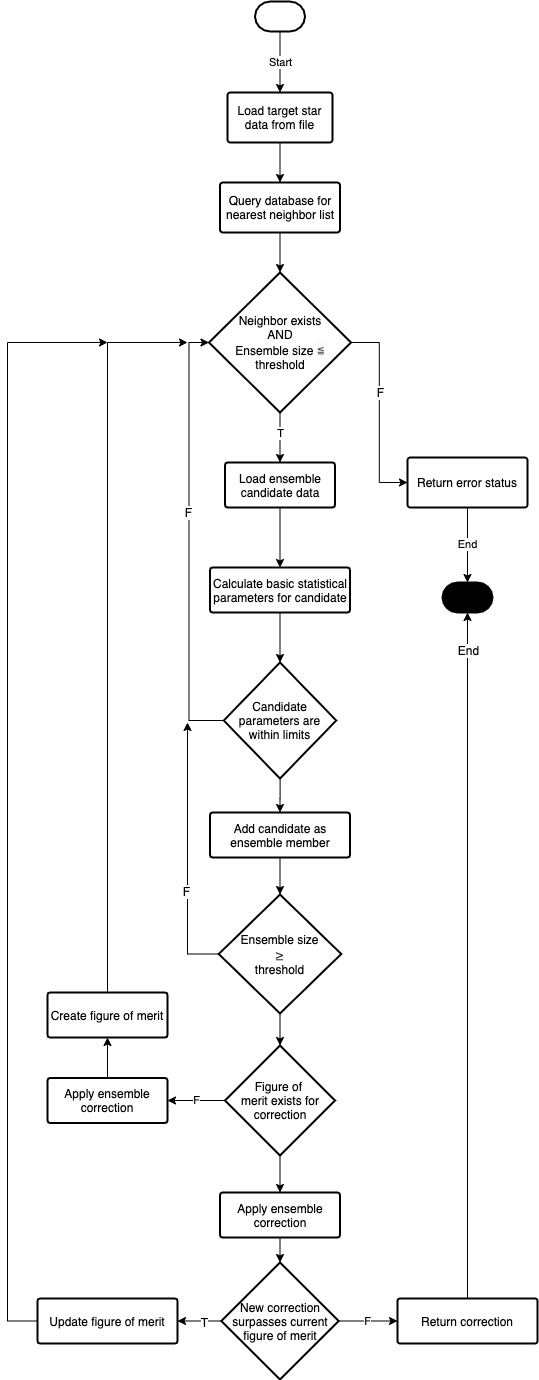}
    \caption{A high-level summary of the Ensemble correction algorithm. The ensemble is initially constructed from light curves for nearby stars that satisfy basic statistical criteria, and ensemble membership is then increased and refined to maximize the light curve figure of merit described in the text. Along with the final correction, the number and identity (TIC IDs) of members of the final ensemble are returned in the light curve header. The end of the algorithm is represented by the black shape.}
    \label{fig:ensflow}
\end{figure}
%%%%%%%%%%%%%%%%%%%%%%%%%%%%%%%%%%%%%%%%%

\item We then loop through the ensemble candidates in order of increasing Euclidean distance from the target, adding each member to the ensemble list if it meets the following criteria, values for which were determined from numerical experiments with targets in TESS Sector~1:
	\begin{enumerate}
	\item The differenced standard deviation of the ensemble candidate does not exceed a maximum value, $\sigma_D < \sigma_{D,max}$. The maximum allowed value $\sigma_{D,max} = 1$, which ensures that the ensemble is not contaminated with very noisy light curves (typically from faint stars). 
	\item The differenced standard deviation of the ensemble candidate is less than ten times that of the target.
	\item We require that the range parameter $R_{var} < 0.4$, to ensure that highly variable stars, such as classical variable stars, are not included in the ensemble. 
	\end{enumerate}
These limiting values for these parameters are recorded in the header to allow for future flexibility. As mentioned above the values for the listed criteria are based on Sector 1 data. Given the improvements of TESS operations from Sector 4 these numbers will possibly be updated in later versions of the pipeline based on new tests. Any such changes will be commented on in the associated data release notes.

The minimum number of stars in the ensemble is 5. Once that number is reached, a fit of the ensemble light curve to the target star is done. We then add the next candidate star to the ensemble and perform the new ensemble fit (see \sref{sec:ens_app}) generating the corrected flux light curve $F_{\rm corr}$. We track the quality of the corrected light curve as a function of the number of ensemble members using the normalized first-order total variation $TV$ \citep{White2017}, or equivalently the geometric length of the time series \citep{Buzasi2015,Prsa2019}, as our figure of merit:
\begin{equation}
    TV = \sum_{i=1}^N \frac{\mid F_{{\rm corr},i} - F_{{\rm corr},i-1} \mid}{M(F_{\rm corr})}\, ,
\end{equation}
where $M$ refers to the median operator. Additional members are added to the ensemble until the quality of the correction, as traced by $TV$, no longer increases; typical resulting ensemble sizes are $\sim$10 stars.

Practically, we have noted that stars with identical TESS magnitudes do not always produce light curves with a similar number of counts, and that these observed offsets are correlated with local background levels, implying that background subtraction is (unsurprisingly) imperfect. We refer to \citetalias{PaperI} for details on the background estimation. For bright stars, minor errors in background estimation produce insignificant offsets, but the impact is more significant for fainter stars. Accordingly, we adjust the background flux level $B$ of each ensemble member to minimize the difference between the flux of the median-filtered ensemble $F_E$ and target $F_t$ light curves, via the statistic $\zeta$:
\begin{equation}\label{eq:ens_back}
\zeta = \sum_{i=0}^{N}\left[ \frac{F_{t,i}}{M(F_t)} - \frac{F_{E,i} + B}{M(F_E +B)} \right] ^2 \, .
\end{equation}
We then apply the optimal offset value $B$, typically a few counts per pixel, to the ensemble light curve. \fref{fig:ens_back} illustrates the results arising from the application of this algorithm to a few members of a typical ensemble.

\begin{figure*}
    \centering
    \includegraphics[width=\textwidth]{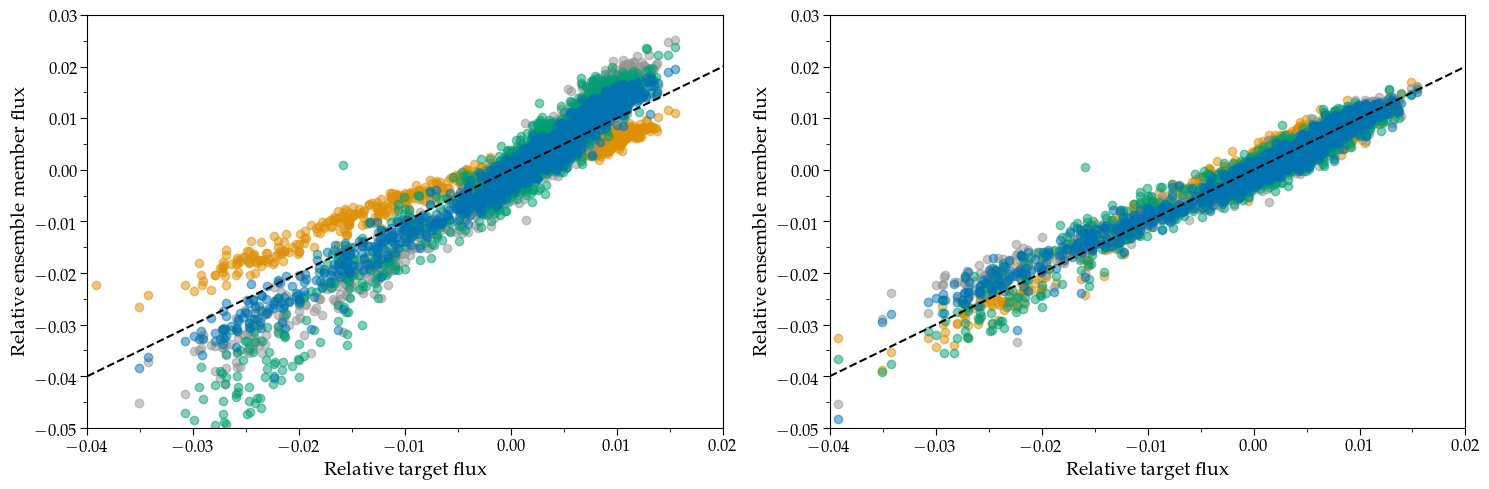}
    \caption{Left: Overlying relative fluxes for 4 photometrically quiet ensemble candidates plotted against the flux for a target star at the same timestamps. Note the varying slopes for several of the ensemble members. Right: The same candidate fluxes after the background correction described in \sref{sec:ens_gen} and \eqref{eq:ens_back}. The dotted lines in both figures illustrated the $1:1$ correlation.}
    \label{fig:ens_back}
\end{figure*}

\subsubsection{Ensemble Application}
\label{sec:ens_app}
To apply the ensemble fit to the target star we remove significant outliers, both instrumental and astrophysical, at each cadence by clipping points which are more than $2\sigma$ from the mean value at that cadence. The ensemble light curves are then median filtered at each time step to produce a single ensemble light curve $F_E$ which can be used for correction purposes. Before application, we allow for a modest flux offset $K$ of $F_E$ through minimization of the scaled Euclidean length of the time series:
\begin{equation}
    F_2  = \frac{\left| \sum_{i=1}^{N} \Delta_i \frac{F_{t,i}}{M(F_E)+K}\right|}{M\left(\frac{F_t}{M(F_E)+K}\right)}
\end{equation}

where $\Delta_i$ represents the $i^{th}$ first difference, where the ``first difference'' is the difference between one time series value and the next \citep{Nason2006}. The resulting offset is applied to $F_E$ and the result is divided into the target light curve to produce a corrected light curve $F_{\rm corr}$. 

\end{itemize}

%%%%%%%%%%%%%%%%%%%%%%%%%%%%%%%%%%%%%%%%%%%%%%%%%%%%%%%%%%%%%%%%%
%%%%%%%%%%%%%%%%%%%%%%%%%%%%%%%%%%%%%%%%%%%%%%%%%%%%%%%%%%%%%%%%%
%%%%%%%%%%%%%%%%%%%%%%%%%%%%%%%%%%%%%%%%%%%%%%%%%%%%%%%%%%%%%%%%%
\section{Performance}\label{sec:per}

\subsection{Photometric performance}
\begin{figure*}
    \centering
    \includegraphics[width=\textwidth]{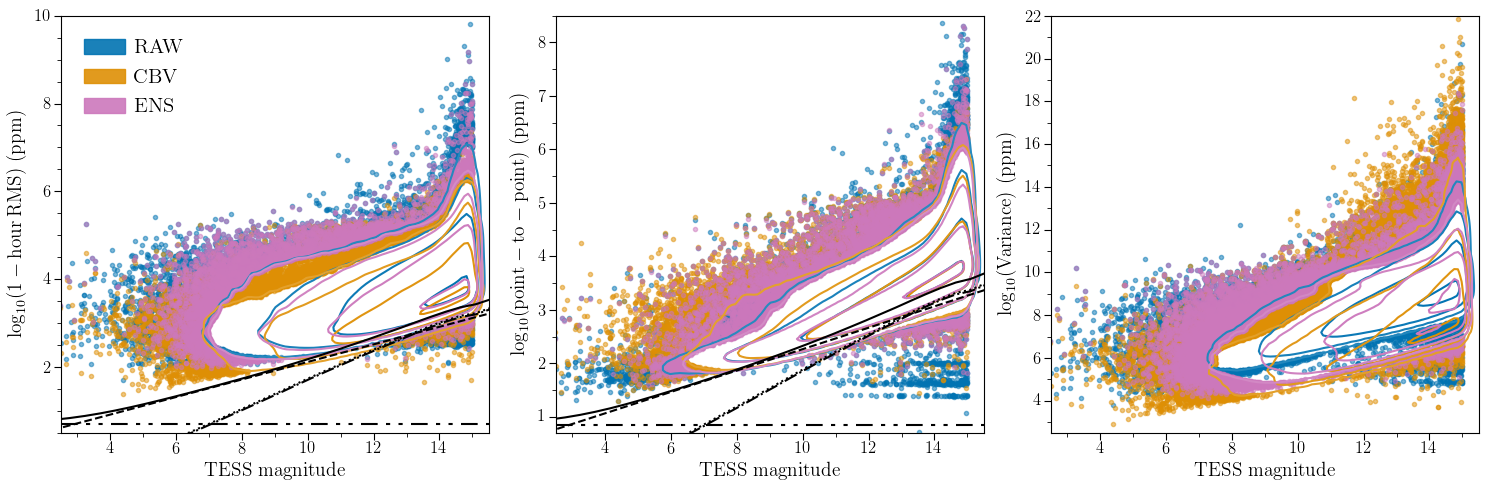}
    \caption{Noise properties (in logarithmic units) as a function of TESS magnitude for stars observed in \LC cadence during Sector 6. The left panel gives the 1-hour root-mean-square (RMS) noise, the middle gives the Point-to-point Median-Differential-Variability (PTP-MDV), and the right panel gives the overall variance of the light curve. The different colors correspond to the different light curve corrections. The distributions are given as contours down to a 1-part per thousand, values outside this level are given as points. The different black lines give the predicted noise estimates following \citet{Sullivan2015} (dashed: shot noise; dotted: read noise; dash-dot: zodiacal noise; dash-dot-dot: systematic noise (here set to 5 ppm); black full: total noise).}
    \label{fig:rms_s6}
\end{figure*}

As is customary in assessing the performance of photometric correction methods, we look first at noise metrics of the resulting light curve. \fref{fig:rms_s6} shows the impact of the systematics correction for Sector 6 \LC cadence targets on the 1-hour root mean square (RMS) variability, the Point-to-point Median-Differential-Variability (PTP-MDV), and the overall variance of the light curve. The differently-colored contours/points refer to values for the different correction methods and the raw photometry.
%%%%%%%%%%%%%%%%%%%%%%%%%%%%%%%%%%%%%%%%
%%%%%%%%%%%%%%%%%%%%%%%%%%%%%%%%%%%%%%%%%

From \fref{fig:rms_s6} we find that (1) the largest reduction happens for the overall variance, followed by the 1-hour RMS and lastly the PTP-MDV where the reduction is minimal; (2) in all cases the reduction is overall largest for the CBV method. The first point indicates that the co-trending most aggressively removes long-term trends, while the point-to-point scatter is largely unaffected, and the second point shows that the CBV method is more aggressive than the Ensemble method. This behavior is as expected and intended, because the shared systematic trends, often originating from variations attributed to the spacecraft, are expected to be somewhat secular and because no priors are used on the fitting of CBV components, increasing the risk of overfitting, especially for stars with a variability longer than the observational baseline. We do, however, see that the lower envelope of the noise metrics follow the expected noise level from \citet{Sullivan2015}. The relatively few cases (values shown as points are all below the 1 per mill level) at high magnitude where the noise fall below the expected level are typically highly contaminated stars (see \citetalias{PaperI}). 

Beyond the agreement between the lower envelope and the expected noise it is difficult to use the noise metrics to evaluate the performance of the co-trending, because our goal is to preserve intrinsic variability not simply to minimize scatter on a given time scale, as is often done for planet searches using, \eg, the 1-hour combined differential photometric precision (CDPP). 

Concerning the number of components $N$ included in the corrections of the CBV method, a dependence is mainly seen on magnitude -- the brighter the star, hence the lesser dominated by photometric noise, the more CBV components are typically included. At magnitudes of ($T_{\rm mag}>12$) the distribution for $N$ is fairly uniform between $3-16$ components. 

We note that it is impossible to provide a single figure of merit which summarises the quality of the corrections performed, as this will depend on the type of star, magnitude, position on the CCD, etc. A user of these light curves will have to make their best judgment of the quality of the corrections conditioned on the type of object being observed.

\subsection{Example light curves}
In order to properly evaluate performance we would need to simulate the light curves with all different types of expected intrinsic variability, covering a range of magnitudes and crowding values, and include known systematic variability covering a range of time scales and relative amplitudes. Such an exercise is beyond the scope of the current paper, but we are planning a comparison for a future paper, with the intent to include all methods currently available for the reduction of TESS photometry.
We can, however, qualitatively evaluate the performance and compare strengths and complementary between the CBV and Ensemble methods by examples of corrected light curves with identifiable intrinsic variability. 

\begin{figure*}
    \centering
    \includegraphics[width=\textwidth]{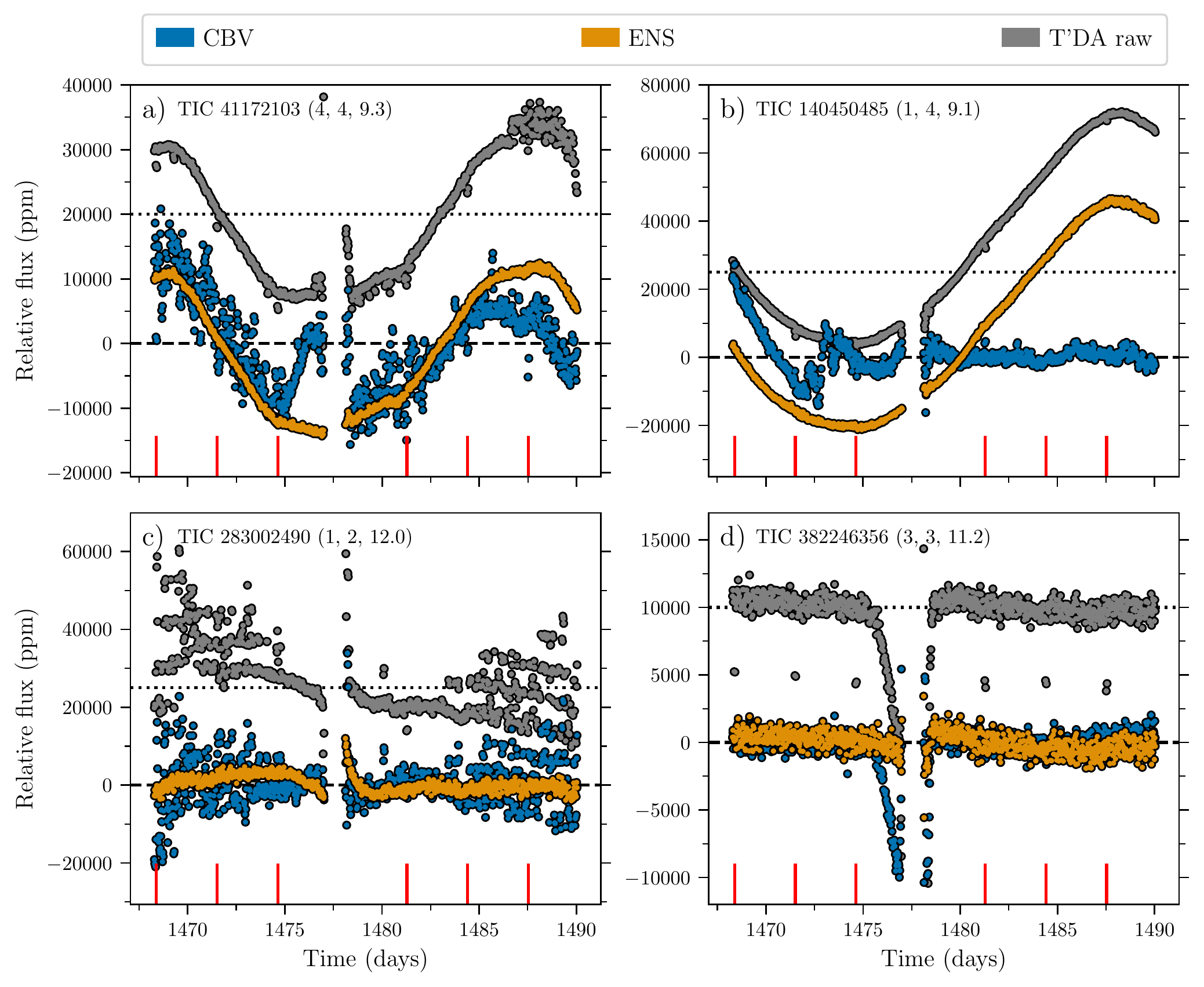}
    \caption{Example light curves for stars observed in \LC during Sector 6, that highlight the particular strengths of the Ensemble (ENS) method. The TIC number of the targets are given in the top of the panels, including in parenthesis the target camera, CCD, and TESS magnitude. The colour of the points/lines indicate the source of data. The raw light curves have been offset vertically from the corrected light curves by the difference between the horizontal dashed and dotted lines. The vertical red lines in the left panels mark the position in time for the reaction wheel desaturations.}
    \label{fig:ex_ens}
\end{figure*}

\begin{figure*}
    \centering
    \includegraphics[width=\textwidth]{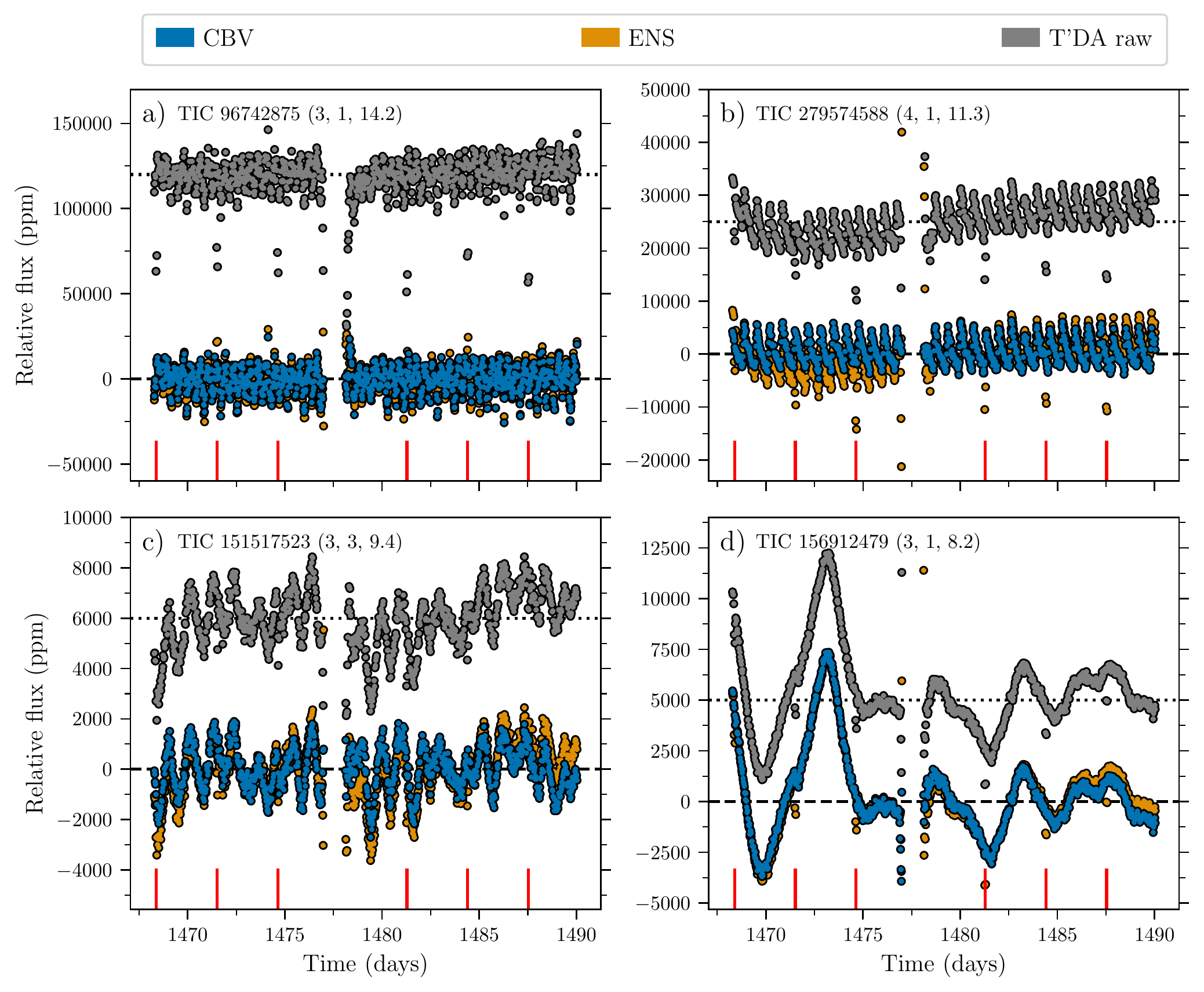}
    \caption{Example light curves for stars observed in \LC during Sector 6, showing cases where both methods are able to preserve the intrinsic stellar variability. The TIC number of the targets are given in the top of the panels, including in parenthesis the target camera, CCD, and TESS magnitude. The colour of the points/lines indicate the source of data. The raw light curves have been offset vertically from the corrected light curves by the difference between the horizontal dashed and dotted lines. The vertical red lines in the left panels mark the position in time for the reaction wheel desaturations.}
    \label{fig:ex_gen}
\end{figure*}

\begin{figure*}
    \centering
    \includegraphics[width=\textwidth]{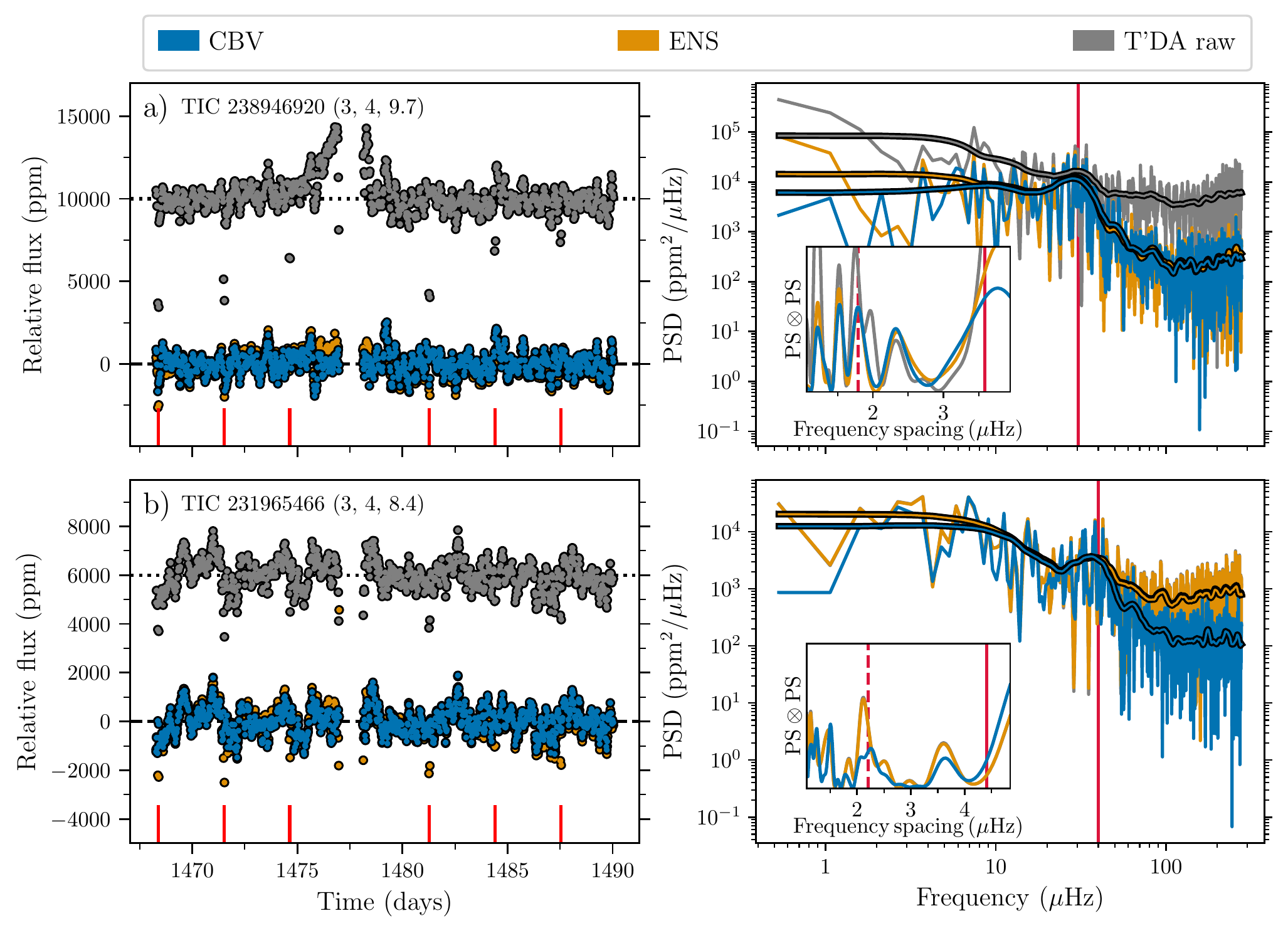}
    \caption{Examples of oscillating red giants observed in \LC data during Sector 6. Left panels show the light curves, while the right panels show the corresponding power density spectra (PDS). The colour of the points/lines indicate the source of data. The TIC number of the targets are indicated in the left panels, including in parenthesis the target camera, CCD, and TESS magnitude. The raw light curves have been offset vertically from the corrected light curves by the difference between the horizontal dashed and dotted lines. The vertical red lines in the left panels mark the position in time for the reaction wheel desaturations. The thick lines in the right panels show the $\rm 15\, \mu Hz$ smoothed version of the PSD. The inserts show the $\rm PS\otimes PS$ of the PSD centred on the identified hump of excess power with \numax indicated by the vertical red line. The corresponding expected \dnu and $\dnu/2$ (dashed line) values following the relation of \citet{huber2011} are marked in the insert. We note that in panel b) the raw and ensemble PDS overlap.}
    \label{fig:ex_osc}
\end{figure*}

\begin{figure}
    \centering
    \includegraphics[width=\columnwidth]{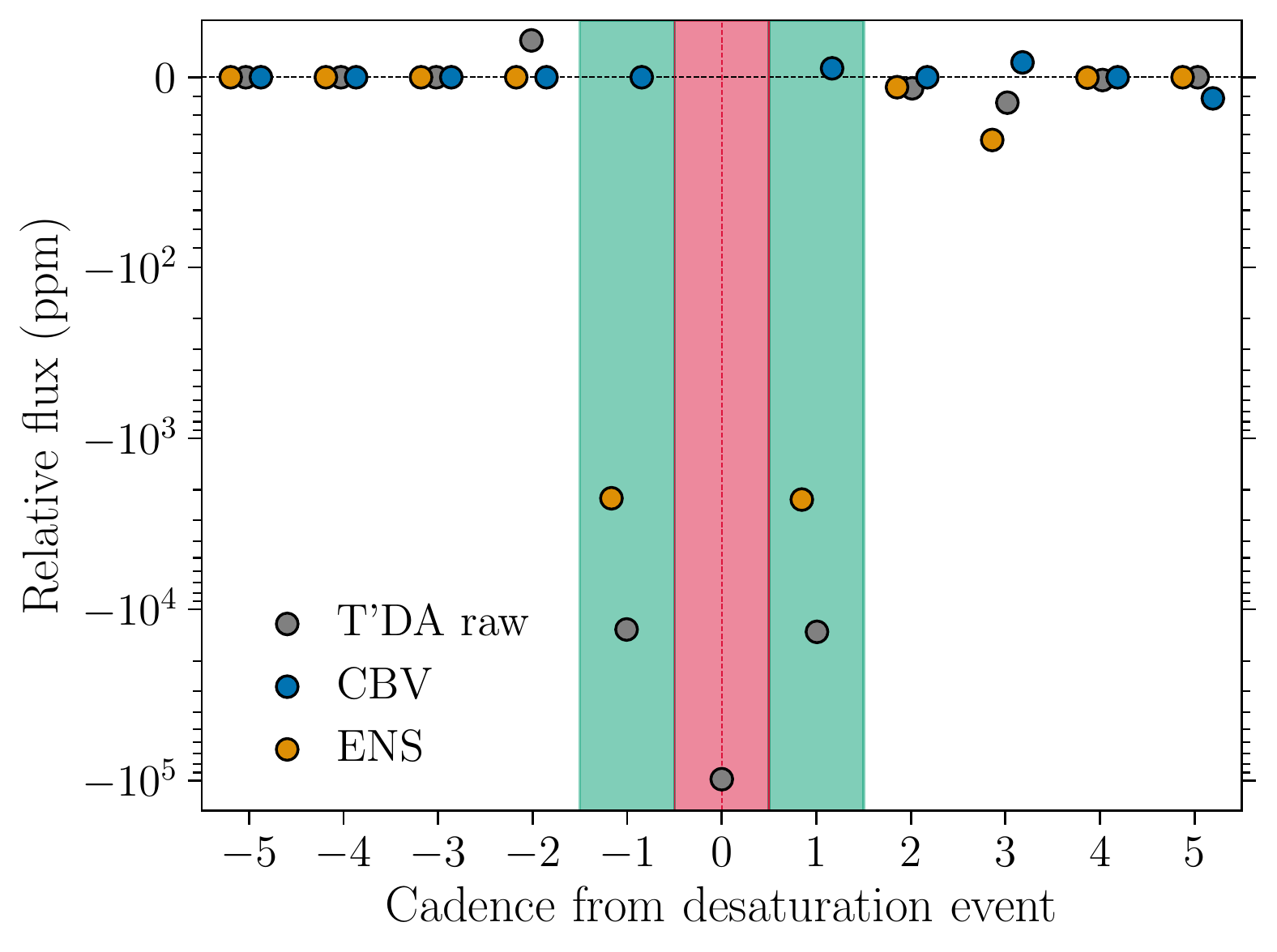}
    \caption{Median-binned relative flux values of 200 random \LC light curves, phase-folded at the period of the reaction wheel desaturation events of ${\sim}3.1250$ days. The colour of the points indicate the data source, and within a cadence the points have been offset horizontally for a better distinction. The light curves have been rectified by a moving median filter with a width of 2 days before the phase folding. The central cadence marked by the red region is captured in the pixel quality flags, hence removed for the CBV and Ensemble (ENS) corrected data. The green regions mark the two cadences flanking the flagged central cadence that are also significantly affected by the desaturation events. The ordinate has a linear scale in the $\pm100$ ppm region and is logarithmic beyond this.   }
    \label{fig:ex_dump}
\end{figure}

Figures~\ref{fig:ex_ens}-\ref{fig:ex_osc} show examples of corrected light curves for different types of variables. These cases were selected from a randomly generated set of 200 targets, selected to uniformly cover the focal plane and a spread in magnitudes.
%%%%%%%%%%%%%%%%%%%%%%%%%%%%%%%%%%%%%%%
%%%%%%%%%%%%%%%%%%%%%%%%%%%%%%%%%%%%%%%%%%%
In \fref{fig:ex_ens} we show four cases highlighting the particular strengths of the Ensemble method -- panels a) and b) give examples of long-period variable (LPV) stars where the Ensemble method is able to preserve the astrophysical signal while also correcting the drops in flux from the momentum wheel desaturation events (see \fref{fig:ex_dump}). In comparison, the CBV method performs poorly for these types of stars from overfitting the variability. Panels a), c) and d) give examples of the local nature of the Ensemble correction method, \ie, building the systematics correction from stars in close proximity on the detector, which enables a correction of the high levels of spatially localized noise seen in these light curves. The local nature of the noise means that it is not well represented in the co-trending basis vectors, and thus the CBV method struggles to properly correct the noise in these cases.

%%%%%%%%%%%%%%%%%%%%%%%%%%%%%%%%%%%%%%%%%
%%%%%%%%%%%%%%%%%%%%%%%%%%%%%%%%%%%%%%%%%%%
\fref{fig:ex_gen} shows examples of stellar variability where both methods are able to preserve the astrophysical signal. However, the CBV method here and in general better corrects the small scale (as compared to the intrinsic variability) systematic trends, most prominently seen in panels b) and c). In general, the CBV method also appears better at correcting the residuals from the reaction wheel desaturation events, which are not captured by the Ensemble method for the light curves in panels b) and d), and the systematics from scattered light near the data downlink gap.

%%%%%%%%%%%%%%%%%%%%%%%%%%%%%%%%%%%%%%%%%%%
%%%%%%%%%%%%%%%%%%%%%%%%%%%%%%%%%%%%%%%%
\fref{fig:ex_osc} shows examples of two oscillating red giants found among the random set of targets. The oscillations are evident from the corrected data of both methods, but the CBV method is seen to perform sightly better in correcting the systematics from scattered light near the data downlink gap and the desaturation events, which remain in the corrected Ensemble data in panel b). We also note that the CBV method has much lower high-frequency noise in the power density spectra compared to the raw data in panel a) and both the raw and ensemble-corrected data in panel b).

Concerning the reaction wheel desaturation events, \fref{fig:ex_dump} shows the representative impact on the light curves, and the correction of this by our methods.
The 200 random \LC light curves were first detrended by a 2-day moving median filter and then phase-folded at a period of ${\sim}3.1250$ days, corresponding to the time between reaction wheel desaturations. The light curve segments before and after the data downlink were treated separately as the phase of the desaturations is shifted at the gap. The folded light curves were then median binned, with a bin-width corresponding to the \LC cadence. The central cadence of the desaturation (marked in red) is impacted the most and this point will be removed using the pixel quality flags (bit 32). However, the neighbouring cadences (marked in green) are also heavily affected, but are not tagged with a quality flag in the photometry. As seen, the CBV method is able to correct for the impact at these cadences, resulting in near-zero median binned flux (note, ordinate in \fref{fig:ex_dump} is linear in the $\pm10$ ppm region). The Ensemble method often leaves a residual, but still reduces the impact by an order of magnitude. 
Because the affected cadences are often properly corrected they are left in the final light curves, but the user should (as always) make sure to assess the quality of the corrections before further analysis. Removing a single cadence on both sides of the flagged desaturations should in any case alleviate issues from uncorrected residuals. We plan in a future version of the pipeline to include, for both methods, a check of the residuals around the flagged desaturations. If the points are found to be outliers because of an insufficient correction they will be flagged as such in the ``\texttt{QUALITY}'' extension of the final data product (see \sref{sec:recommend} and \citetalias{PaperI}, Table~1). The inclusion of such an additional check will be noted in the data release notes associated to the given version of the pipeline.   
%%%%%%%%%%%%%%%%%%%%%%%%%%%%%%%%%%%%%%%%
%%%%%%%%%%%%%%%%%%%%%%%%%%%%%%%%%%%%%%%%

\begin{figure*}
    \centering
    \includegraphics[width=0.905\textwidth]{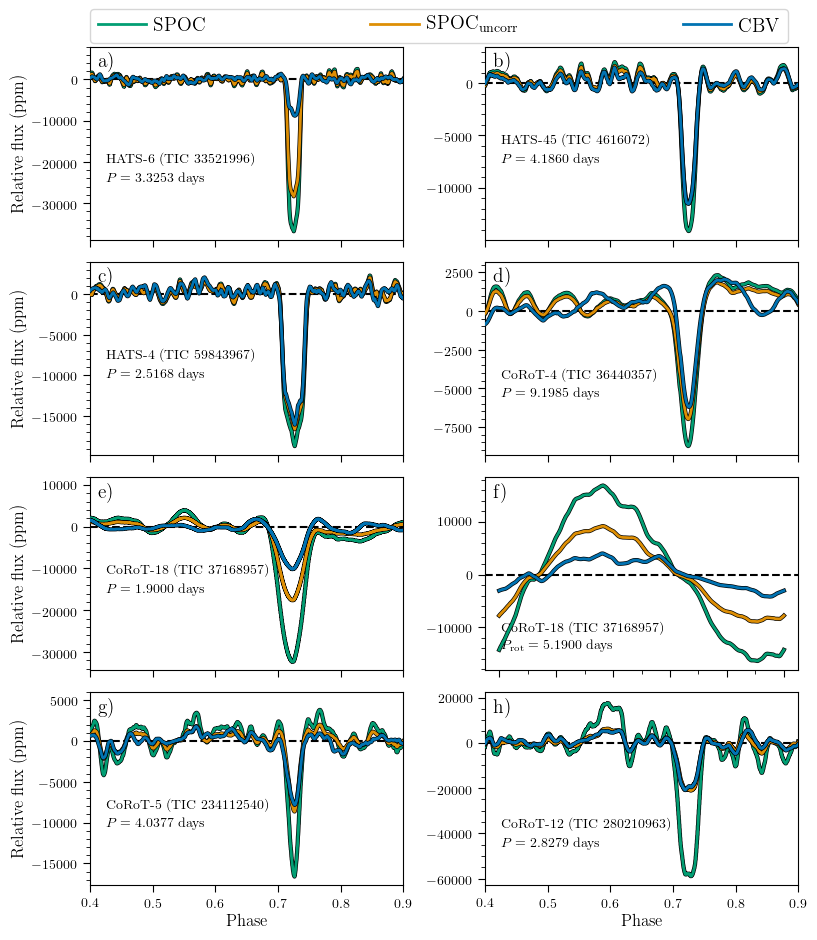}
    \caption{Examples of phase-folded systematics-corrected light curves for known exoplanets observed during Sector~6. The light curves have been smoothed by a moving median filter with a width corresponding to $5\%$ of the transit duration (for panel f the filter width is $1\%$ of the stellar rotation period). The names of the individual systems along with the periods used are indicated in the panels. The colour of the lines indicate the different corrections from the CBV method and the \texttt{PDCSAP} from SPOC. The $\rm SPOC_{uncorr}$ gives the version of the SPOC data after reverting the correction for crowding and flux fraction captured in the aperture. Panels e) and f) both show data for CoRoT-18, displaying the transit in panel e) and the rotational modulation in panel f) \citep[see][]{corot18-2011,Raetz2019}.} 
    \label{fig:planet}
\end{figure*}

In \fref{fig:planet} we show examples of light curves for known exoplanet systems observed during Sector 6, including the systems CoRoT-4 \citep{corot4}, CoRoT-5, CoRoT-12, CoRoT-18 \citep{corot18-2011}, HATS-4 \citep{HATS4}, HATS-6 \citep{HATS6}, and HATS-45 \citep{HATS45}.
For these \SC targets we show in addition to the CBV light curves the ones produced by SPOC (raw light curves are \texttt{SAP} -- \textit{simple aperture photometry}; corrected light curves are \texttt{PDCSAP} -- \textit{presearch data conditioning SAP}). We omit showing light curves from the \code{eleanor} pipeline \citep[][]{eleanor} because the flux extraction and correction here requires user input on a star by star basis, and data from the method of \citet{Oelkers2018} is currently unavailable for Sector 6.
%%%%%%%%%%%%%%%%%%%%%%%%%%%%%%%%%%%%%%%%
%%%%%%%%%%%%%%%%%%%%%%%%%%%%%%%%%%%%%%%%%
\begin{figure*}
    \centering
    \includegraphics[width=\textwidth]{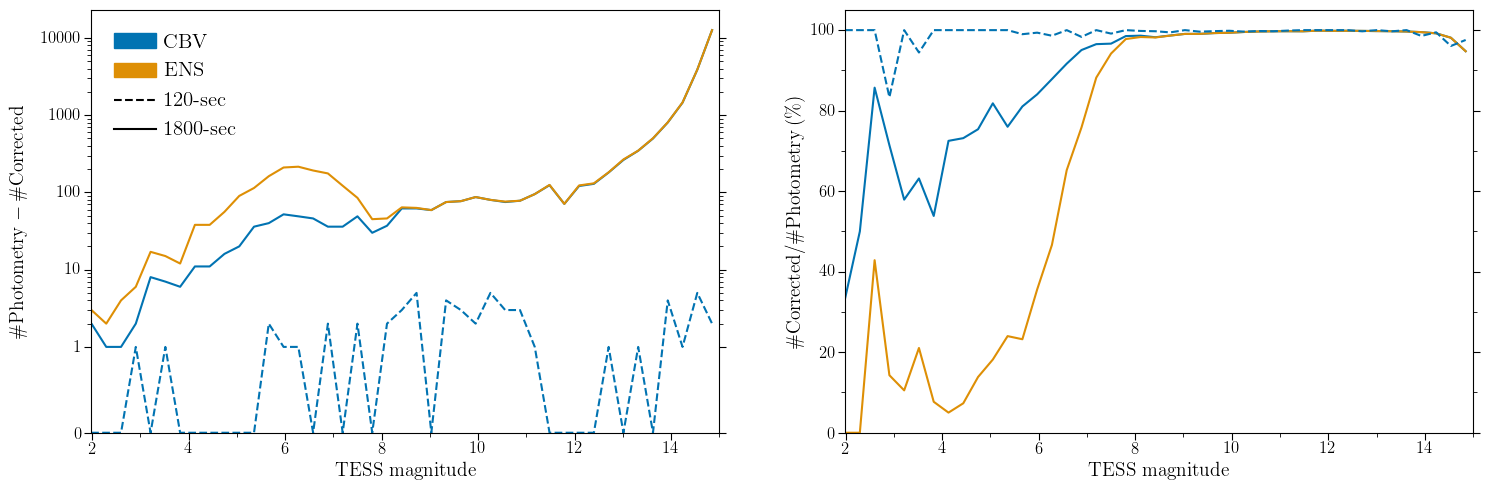}
    \caption{Left: Difference between number of corrected targets and number of targets with available photometry as a function of TESS magnitude. Right: Percentage ratio of corrected light curves as a function of TESS magnitude. }
    \label{fig:numbers_comp}
\end{figure*}

We see that in all cases the planetary transits are clearly visible when phase-folding the data at the planetary orbital period\footnote{obtained from the TOI-catalog \url{https://tev.mit.edu/data/collection/193/}}. When comparing the transits from the CBV method with SPOC, we see that the transit depth from SPOC data often is significantly larger. In most cases this is a consequence of the correction of SPOC \texttt{SAP} (and therefore also \texttt{PDCSAP}) data for both an estimated crowding and the target star flux fraction captured in the optimum aperture \citep[see][]{pdc1}. If we revert the applied corrections\footnote{the uncorrected flux $F_0$ is obtained from the \texttt{PDCSAP} flux $F$ as $F_0 = f(F + M(F)(1/c - 1))$ where $c$ is the crowding factor (\texttt{CROWDSAP}), $f$ is the flux fraction (\texttt{FLFRCSAP}), and $M$ is the median operator.} to SPOC data we generally see an excellent agreement in the transit shapes, which highlights the validity of using T'DA data for exoplanet searches and analysis -- and importantly, T'DA also offers data for all identified \LC FFI targets. 
Remaining differences in the transit shapes, specifically the depth, can be traced to the raw photometry from different aperture sizes (\citetalias{PaperI}), most prominently seen for HATS-6 (panel a) and CoRoT-18 (panel e). It is clear, however, that with a correction for crowding and fractional flux contained in the aperture, which should be applied before an analysis of the transit, most of the differences between the CBV and the SPOC corrected data should be remedied. We leave this task to the expert user -- as seen from \fref{fig:planet} the effect of the correction by SPOC can be quite significant, and appear in some cases overestimated based on a comparison with transit depths from the literature. We see also that for CoRoT-18 (panel f) we are able to preserve the rotational modulation of the star in addition to the exoplanet transit (panel e). We note that all the know exoplanet hosts shown are relatively faint ($T_{\rm mag}>12.75$) and crowded.

\subsection{Correction yield}
It is also worth comparing the numbers of corrected light curves for the Ensemble and CBV methods. Due to the architecture of the Ensemble code, there is a requirement for a certain number of stars in the vicinity of the target star with similar magnitudes. As the target star gets brighter it becomes increasingly difficult to find suitable nearby comparison stars for the ensemble. In \fref{fig:numbers_comp} we compare the number of corrected light curves with the number of available light curves as a function of magnitude. As seen, the Ensemble method will for \LC targets generally return only a high percentage of the stars with TESS magnitude $\gtrsim 7-8$. The numbers do of course differ between campaigns from variations in field density etc., but the user of Ensemble light curves should not be surprised if a corrected light curve is unavailable for relatively bright targets.

\subsection{Processing time}
\begin{figure}
    \centering
    \includegraphics[width=\columnwidth]{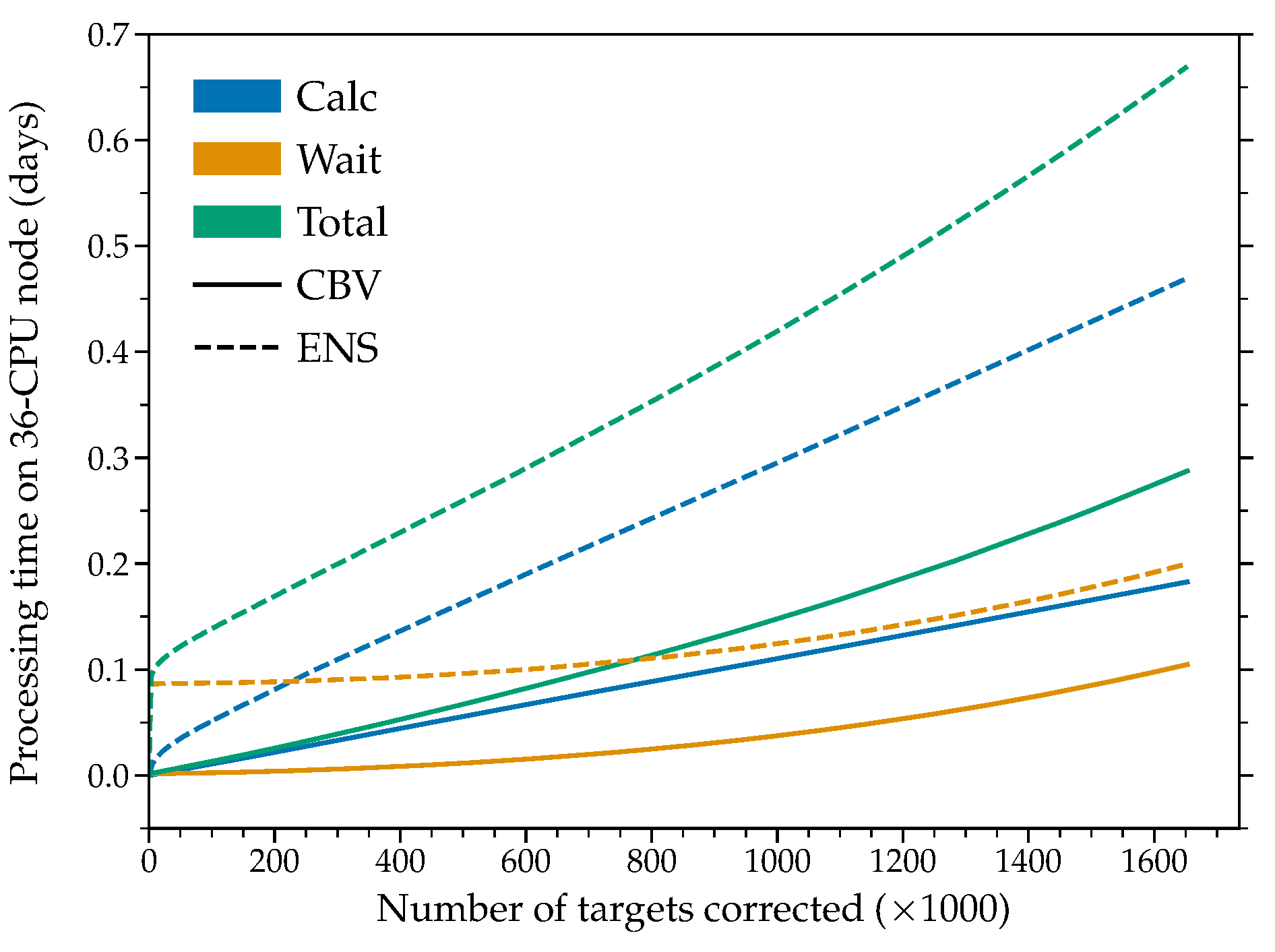}
    \caption{Cumulative processing time in days for \LC cadence targets from Sector 6 with our current setup of a 36-CPUcore node. The different colors refer to the different components of the total (green) processing time, namely the actual calculation time (blue) and the idle wait time (yellow) for writing results to a shared output file. The full line gives the values for the CBV method and the dashed gives the corresponding for the Ensemble method (ENS).}
    \label{fig:processtime}
\end{figure}

In terms of calculation time the co-trending of targets with the CBV approach proceeds at a pace of ${\sim}0.34$ seconds per star, while the Ensemble method takes ${\sim}0.88$ seconds per star. For the CBV approach, this calculation time assumes that the CBVs have been generated, a process that in itself takes of the order ${\sim}0.5-1$ days. \fref{fig:processtime} shows the total processing time for all \LC cadence targets in Sector 6 with our current setup with a single 36-CPUcore node on the Grendel-Slurm cluster at the Center for Scientific Computing Aarhus (CSCAA\footnote{\url{http://www.cscaa.dk/grendel-s/overview/}}).
%%%%%%%%%%%%%%%%%%%%%%%%%%%%%%%%%%%%%%%%
%%%%%%%%%%%%%%%%%%%%%%%%%%%%%%%%%%%%%%%%%
The total processing time of ${\sim}1.650.000$ stars amounts to ${\sim}0.3$ and ${\sim}0.7$ days for the CBV and Ensemble methods respectively. A secondary time component entering the total processing time comes from idle time where a process waits to start because output is being written by other processes to a file shared amongst the processes on the node. While further optimisations are being pursued the processing is at present fast enough to keep up with the extraction of raw photometry, and the processing can be parallelized over several nodes if required.

\subsection{Light curve recommendations}\label{sec:recommend}
Detailed recommendations on best practices are still being investigated, but in general we suggest that users consult corrected data from both methods to see if they are compatible. In cases where there are large differences the underlying stellar variability will typically be of the long-period kind, and in these cases then Ensemble data should generally be preferred. When the corrected light curves are compatible (difference below or at the level of the variability range of the raw light curve) the CBV method will generally do a better correction of the light curve systematics, including the reaction wheel momentum desaturations (\fref{fig:ex_dump}) and the strong systematics from scattered light around the data downlink gaps.
We recommend that users always consult the quality flags accompanying the raw and corrected light curves, and the release notes associated with the data. As detailed in \citetalias{PaperI} the corrected light curves contain both a column of ``\code{PIXEL$\_$QUALITY}'', which gives the quality flag provided by the TESS team (see the  \href{https://outerspace.stsci.edu/display/TESS/2.0+-+Data+Product+Overview}{TESS Archive Manual}), and of ``\code{QUALITY}'' which gives the quality flags set by the TASOC pipeline (see \citetalias{PaperI}, Table~1). We also encourage users to specifically check the corrected light curves around the downlink gaps -- while the strong systematics typically found here from the scattered light of the Earth are properly corrected in many cases, especially for the CBV method, residuals can occur and could hamper a subsequent analysis if not removed or detrended from the light curve. We note that the corrected light curves are not gap-filled. It is up to the user of the data to decide if gap-filling/inpainting of the data is useful for the specific science case and in that case what method should be adopted \citep[see, \eg,][]{rafa2014,2018A&A...614A..40P}.
%%%%%%%%%%%%%%%%%%%%%%%%%%%%%%%%%%%%%%%%%%%%%%%%%%%%%%%%%%%%%%%%%
%%%%%%%%%%%%%%%%%%%%%%%%%%%%%%%%%%%%%%%%%%%%%%%%%%%%%%%%%%%%%%%%%
%%%%%%%%%%%%%%%%%%%%%%%%%%%%%%%%%%%%%%%%%%%%%%%%%%%%%%%%%%%%%%%%%

\section{Data format and access}\label{sec:format}

We refer to \citetalias{PaperI} for the general overview of the data formatting, but describe here the details specific to systematics-corrected light curves.

Similar to the raw photometry, the foremost data format for corrected light curves is FITS (Flexible Image Transport System\footnote{\url{https://fits.gsfc.nasa.gov/fits\_standard.html}}), and is provided in a compressed Gzip format.
A FITS light curve file produced by T'DA for a corrected target will be named following the structure:
\begin{center}\footnotesize
    \code{tess$\{$TIC~ID$\}$-s$\{$sector$\}$-$\{$camera$\}$-$\{$ccd$\}$-c$\{$cadence$\}$\\-dr$\{$data~release$\}$-v$\{$version$\}$-tasoc-$\{$method$\}\_$lc.fits.gz}
\end{center}
The ``TIC ID'' (TESS Input Catalog identifier) of the star is zero (pre-)padded to 11 digits, the ``sector'' is zero (pre-)padded to 3 digits, the ``cadence'' is in seconds and zero (pre-)padded to 4 digits, the ``data release'' is zero (pre-)padded to 2 digits and refers to the official release of the data from the mission, the ``version'' is zero (pre-)padded to 2 digits and refers to the TASOC data release (counting from 1). The file-naming is thus identical to that of the raw photometry, with the exception of the ``method'', which refers to the co-trending method used. The ``method'' can either be ``cbv'' for the CBV approach (\sref{sec:cbv}) or ``ens'' for the Ensemble approach (\sref{sec:ens}).

When accessing the data files via MAST\footnoteref{note_mast} (\dataset[https://doi.org/10.17909/t9-4smn-dx89]{https://doi.org/10.17909/t9-4smn-dx89}) as a `high level science product'' the filename will be slightly different --- we refer to MAST for a description of the naming convention.

Each light curve FITS file has four extensions: a ``\code{PRIMARY}'' header with general information on the star and the observations; a ``\code{LIGHTCURVE}'' table with time, raw flux, corrected flux, etc.; a ``\code{SUMIMAGE}'' with an image given by the time-averaged pixel data; and an ``\code{APERTURE}'' image. The information provided in the FITS file is intended to mimic that provided in the official TESS products -- please consult the ``TESS Science Data Products Description''\footnote{\url{https://archive.stsci.edu/missions/tess/doc/EXP-TESS-ARC-ICD-TM-0014.pdf}} for more information.

The header of the ``\code{LIGHTCURVE}'' extension will provide some details on the cotrending. In addition to the filename, the key ``\code{CORRMET}'' will provide the adopted correction method, while ``\code{CORRVER}'' will refer to the version used of the correction module, according to the version tag of the code on GitHub.

In the case of a CBV correction (\sref{sec:cbv}) the header will include values for: the weights of the fitted CBVs as ``\code{CBV$\_$c\#}'' (normal) and ``\code{CBVS$\_$c\#}'' (spike), where \code{\#} refers to the number of the CBV; the ``\code{CBV$\_$AREA}'' the star belongs to; the number of fitted CBVs (``\code{CBV$\_$COMP}''); if BIC was used to decide on the best number of CBVs (``\code{CBV$\_$BIC}''); the number of CBVs allowed in the fit (``\code{CBV$\_$MAX}''), \ie, the number of CBVs passing the SNR-test (\sref{sec:cbv}). Information will also be provided on the method used to fit the CBVs (``\code{CBV$\_$MET}'').

In the case of an Ensemble correction (\sref{sec:ens}) the ``\code{LIGHTCURVE}'' extension header will include values for the number of targets included in the ensemble  (``\code{ENS$\_$NUM}''). An additional ``\code{ENSEMBLE}'' extension will list the TIC IDs of the ensemble stars, and the associated value for the background correction parameter $B$ (\eqref{eq:ens_back} as ``\code{BZETA}''). We refer to Appendix~\ref{sec:appa} for examples of the full headers of the \code{LIGHTCURVE} and \code{ENSEMBLE} extensions.

\subsection{CBV files}

The CBVs generated for the processing will also be made available for full reproducibility of the co-trending.
The CBVs will be available in FITS format and named following the structure:
\begin{center}\footnotesize
    \code{tess-s$\{$sector$\}$-c$\{$cadence$\}$-a$\{$area$\}$
    -v$\{$version$\}$-tasoc$\_$cbv.fits}
\end{center}
The ``sector'' is zero (pre-)padded to 4 digits, the ``cadence'' is in seconds and zero (pre-)padded to 4 digits, the ``area'' is the 3 digit cbv-area introduced in \sref{sec:cbv_gen}, and the ``version'' is zero (pre-)padded to 2 digits and refers to the TASOC data release (counting from 1).
The file currently has two extensions, one for the ``normal'' CBVs named \code{CBV.SINGLE-SCALE.$\{$area$\}$}, and one for the ``spike'' CBVs named \code{CBV.SPIKE.$\{$area$\}$}. Each extension gives a column with time and cadence number, and then the respective CBV vectors for the area. If the pipeline, as envisioned, is later extended to use multi-scale CBVs these will be added in a separate extension. 

As mentioned in \sref{sec:cbv_fit}, \SC cadence targets are fitted using interpolated \LC cadence CBVs, hence CBV files will only be available at a \LC cadence.

\subsection{Availability and Data policy}
Data for processed targets can first and foremost be obtained via the TASOC\footnoteref{note_tasoc} database.
The TASOC data releases are described, including the data release notes, under the ``Data Releases'' tab. Individual target search is accessed via the ``Data Search'' tab, and here specifying the ``Data types'' to include ``TASOC Lightcurves''.

Data will also be available at MAST\footnoteref{note_mast} as a High level Science Product (\dataset[HLSP]{https://dx.doi.org/10.17909/t9-4smn-dx89}\footnote{\url{https://archive.stsci.edu/hlsp/}}). We note that in addition to a traditional search on MAST, the data can also be accessed programmatically using the MAST search option in \code{Astroquery}\footnote{\url{https://astroquery.readthedocs.io/en/latest/mast/mast.html}} and is directly accessible via the \code{Lightkurve} \citep{lightkurve} package using the \texttt{author="TASOC"} input parameter when searching for light curves.

%%%%%%%%%%%%%%%%%%%%%%%%%%%%%%%%%%%%%%%%%%%%%%%%%%%%%%%%%%%%%%%%%
%%%%%%%%%%%%%%%%%%%%%%%%%%%%%%%%%%%%%%%%%%%%%%%%%%%%%%%%%%%%%%%%%
%%%%%%%%%%%%%%%%%%%%%%%%%%%%%%%%%%%%%%%%%%%%%%%%%%%%%%%%%%%%%%%%%
\section{Conclusions and outlook}\label{sec:out}

We have presented a module for the TASOC pipeline for producing systematic corrected light curves from FFIs and TPF data. The two methods currently included, the CBV (\sref{sec:cbv}) and the Ensemble methods (\sref{sec:ens}), allows for the processing of stars with a wide range of variability time scales and amplitudes. The Ensemble method is found to best preserve the astrophysical signal for stars with long-period variability, while the CBV method is better in general at removing the secular and often small-scale systematic trends from the TESS orbit, and imprints from the regular reaction wheel desaturation events. 

We find that the processing of photometry from a full sector, typically comprising ${\sim}1.6$ million stars, can be completed within a few days with our current setup at the Centre for Scientific Computing, Aarhus. 

We have several envisioned improvements and extensions to the current pipeline. The goal is to have a pipeline with only one method implemented that can properly treat all cases of variability. As seen in \sref{sec:per} the advantage of having both the CBV and the Ensemble method is their strengths for different timescale of variability, where the CBV method for long period variability has a tendency to over-fit. This behaviour is expected given the lack of constraints on the CBV scaling coefficients. 

\begin{figure}
    \centering
    \includegraphics[width=\columnwidth]{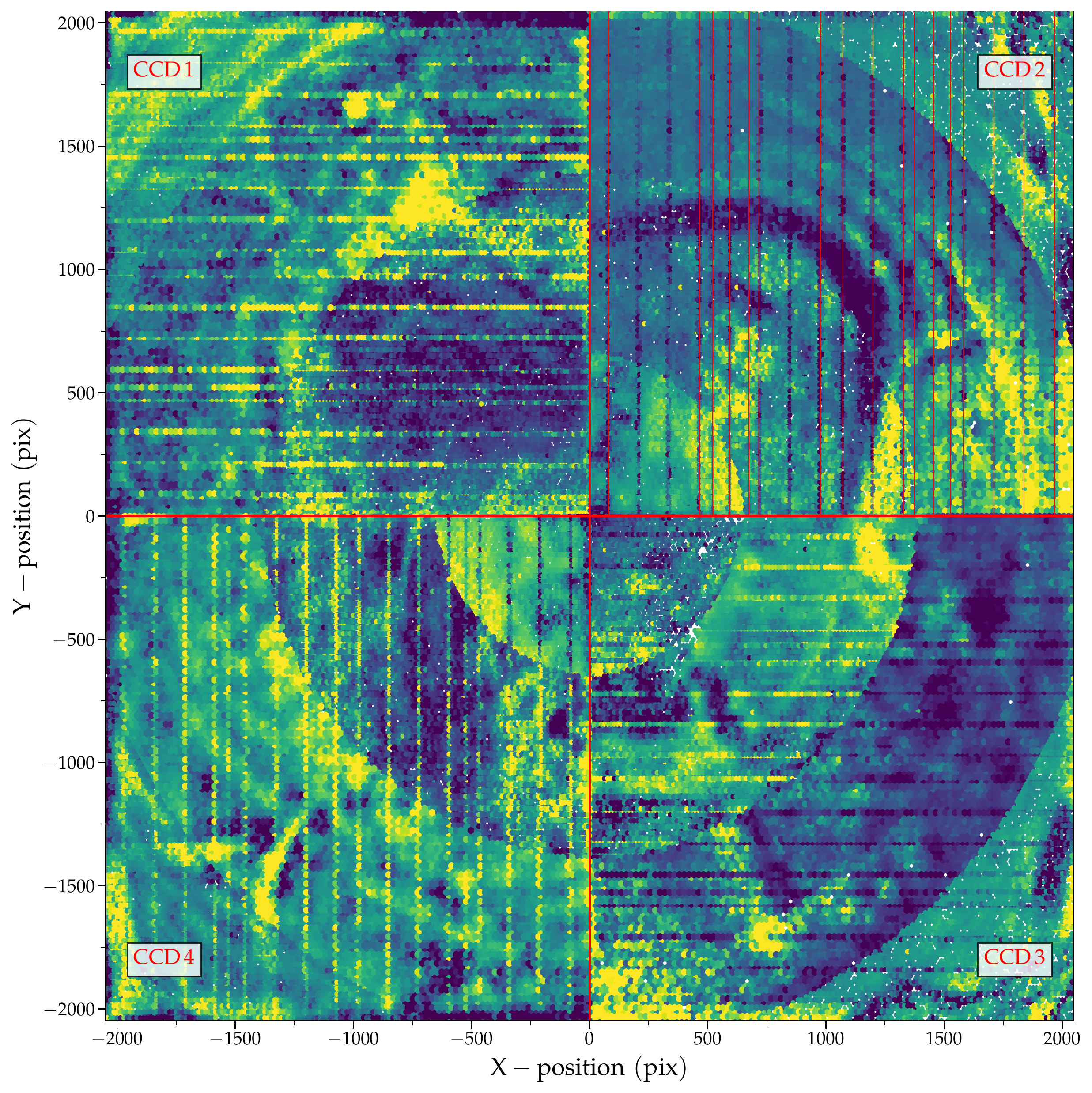}
    \caption{Median binned map of scaling coefficients from the fit of CBV number 1 ($i=1$ in \eqref{eq:fit}) for targets in sector 6 camera 4. The pixel coordinates are given relative to the camera center. Note that the variability captured by the specific CBV can vary for the different CBV areas (see \fref{fig:ccd}) -- this is clearly visible for CCDs 3-4 where the segmentation into co-centric areas is obvious. The vertical/horizontal stripes show the positions of the conducting straps, whose known positions are marked by vertical red lines in CCD 2.}
    \label{fig:wei}
\end{figure}

However, if constraints could be placed on the coefficients based on the cohort of nearby stars in the form of a prior, it should be possible to gain the benefits of the Ensemble method (which utilizes information from nearby stars) but still using orthonormal basis vectors. To indicate the potential of using shared information for the construction of fitting priors we show in \fref{fig:wei} a median-binned map of the scaling coefficients for CBV number 1 ($i=1$ in \eqref{eq:fit}) for targets in Sector 6 camera 4. For the run used to generate this plot all possible CBVs were included in the fits, \ie, the BIC was not used for selecting the number of CBVs to include. As seen there are clear structures in the scaling coefficients within a given CBV area, and in some cases across different areas if the CBV number 1 happen to capture the same component of the variability. One also sees some apparent residuals from the corner-glows (see \citetalias{PaperI}) that are captured by the CBVs. In the current testing we build for a given star and a given CBV a prior by a weighted kernel density estimate (KDE) for the scaling coefficients. The weighting is given by the distance between the target star and those contributing to the prior in the space of position on the CCD and in TESS magnitude. This approach is very similar to that adopted in the \kp mission for the PDC-MAP data product \citep{kepdatc8}. We are still in the testing phase of the implementation of priors. 
Similar to \citet{kepdatc8} we also plan to add a multi-scale component to the CBVs, where sets of CBVs are created for different scales of variability.

An important addition will also be a potential iteration of the co-trending with the variability classification of the pipeline (\citetalias{PaperIII}). If the stellar type is confidently identified from the classification this can be used to guide which correction method is best to use, or possibly a weighting of different scales of CBVs. Tests of such an iteration will be started once the current versions of the correction and classification components of the pipeline are running in a steady state.

We are in the process of improving the Ensemble method for dealing with \SC data. We note, however, that the \SC targets for asteroseismology are unlikely to benefit from adopting the Ensemble approach over the CBV one, as these are generally stochastic solar-like oscillators.  

As mentioned in \sref{sec:per} we are planning a larger comparison of TESS FFI reduction methods currently available. This should serve as a guide to the community on the use of different publicly available data for specific use cases.
%%%%%%%%%%%%%%%%%%%%%%%%%%%%%%%%%%%%%%%%
%%%%%%%%%%%%%%%%%%%%%%%%%%%%%%%%%%%%%%%%%
%%%%%%%%%%%%%%%%%%%%%%%%%%%%%%%%%%%%%%%%%%%%%%%%%%%%%%%%%%%%%%%%%
%%%%%%%%%%%%%%%%%%%%%%%%%%%%%%%%%%%%%%%%%%%%%%%%%%%%%%%%%%%%%%%%%
%%%%%%%%%%%%%%%%%%%%%%%%%%%%%%%%%%%%%%%%%%%%%%%%%%%%%%%%%%%%%%%%%

%%%%%%%%%%%%%%%%%%%%%%%%%%%%%%%%%%%%%%%%%%%%%%%%%%%%%%%%%%%%%%%%%
%%%%%%%%%%%%%%%%%%%%%%%%%%%%%%%%%%%%%%%%%%%%%%%%%%%%%%%%%%%%%%%%%
%%%%%%%%%%%%%%%%%%%%%%%%%%%%%%%%%%%%%%%%%%%%%%%%%%%%%%%%%%%%%%%%%
\acknowledgments
We thank the two anonymous referees for their comments and suggestions which helped improve the paper.
Funding for the Stellar Astrophysics Centre is provided by The Danish National Research Foundation (Grant agreement no.: DNRF106). 
The numerical results presented in this work were obtained at the Centre for Scientific Computing, Aarhus\footnote{\url{http://phys.au.dk/forskning/cscaa/}}.
This research was partially conducted during the Exostar19 program at the Kavli Institute for Theoretical Physics at UC Santa Barbara, which was supported in part by the National Science Foundation under Grant No. NSF PHY-1748958.
MNL and RH acknowledge the support from the ESA PRODEX programme.
DLB, LC, and DH acknowledge support from the TESS GI Program under Grant 80NSSC18K1585.
LC acknowledges support from the European Research Council (ERC) under the European Union's Horizon 2020 research and innovation programme (CartographY GA. 804752).
OJH acknowledges the support of the UK Science and Technology Facilities Council (STFC).
FP acknowledges support from fellowship PD/BD/135227/2017 funded by FCT - Funda\c{c}\~{a}o para a Ci\^{e}ncia e Tecnologia (Portugal) and POPH/FSE - Programa Operacional Potencial Humano (EC).
DH acknowledges support from the Alfred~P.~Sloan Foundation.
The research leading to these results has received funding from the KU Leuven Research Council (grant C16/18/005: PARADISE), and from the Research Foundation Flanders (FWO) by means of  a junior postdoctoral fellowship with grant agreement No. 12ZB620N.
This paper includes data collected by the TESS mission, which are publicly available from the Barbara A. Mikulski Archive for Space Telescopes (MAST). Funding for the TESS mission is provided by NASA's Science Mission directorate. Funding for the TESS Asteroseismic Science Operations Centre is provided by the Danish National Research Foundation (Grant agreement no.: DNRF106), ESA PRODEX (PEA 4000119301) and Stellar Astrophysics Centre (SAC) at Aarhus University. We thank the TESS team and staff and TASC/TASOC for their support of the present work.

\facilities{TESS, MAST (TESS)}

\software{Astropy \citep{astropy:2013, astropy:2018}, scikit-learn \citep{scikit-learn}, NumPy \citep{harris2020array}, SciPy \citep{Scipy}, Matplotlib \citep{matplot}, \href{https://bottleneck.readthedocs.io}{Bottleneck}, Lightkurve \citep{2018ascl.soft12013L}, Seaborn \citep{waskom2020seaborn}}

\newpage
\clearpage

\appendix
\vspace{5cm}
\onecolumngrid
\section{FITS headers of output light curves}\label{sec:appa}
We here include an example of the full headers of the \code{LIGHTCURVE} extension of typical FITS light curve file from both the CBV and Ensemble methods, and the header of the \code{ENSEMBLE} extension for the Ensemble method (see \sref{sec:format}).

\subsection*{CBV HDU-1 (LIGHTCURVE)}
\begin{verbatim}
XTENSION= 'BINTABLE'           / binary table extension
BITPIX  =                    8 / array data type
NAXIS   =                    2 / number of array dimensions
NAXIS1  =                   96 / length of dimension 1
NAXIS2  =                  993 / length of dimension 2
PCOUNT  =                    0 / number of group parameters
GCOUNT  =                    1 / number of groups
TFIELDS =                   14 / number of table fields
INHERIT =                    T / inherit the primary header
EXTNAME = 'LIGHTCURVE'         / extension name
TIMEREF = 'SOLARSYSTEM'        / barycentric correction applied to times
TIMESYS = 'TDB     '           / time system is Barycentric Dynamical Time (TDB)
BJDREFI =              2457000 / integer part of BTJD reference date
BJDREFF =                  0.0 / fraction of the day in BTJD reference date
TIMEUNIT= 'd       '           / time unit for TIME, TSTART and TSTOP
TSTART  =    1468.276639141453 / observation start time in BTJD
TSTOP   =     1490.04694595576 / observation stop time in BTJD
DATE-OBS= '2018-12-15T18:37:12.438' / TSTART as UTC calendar date
DATE-END= '2019-01-06T13:06:26.946' / TSTOP as UTC calendar date
MJD-BEG =    58467.77663914146 / observation start time in MJD
MJD-END =    58489.54694595576 / observation start time in MJD
TELAPSE =    21.77030681430688 / [d] TSTOP - TSTART
LIVETIME=    4.310520749232762 / [d] TELAPSE multiplied by DEADC
DEADC   =                0.198 / deadtime correction
EXPOSURE=    4.310520749232762 / [d] time on source
XPOSURE =                356.4 / [s] Duration of exposure
TIMEPIXR=                  0.5 / bin time beginning=0 middle=0.5 end=1
TIMEDEL =  0.02083333333333333 / [d] time resolution of data
INT_TIME=                 1.98 / [s] photon accumulation time per frame
READTIME=                 0.02 / [s] readout time per frame
FRAMETIM=                  2.0 / [s] frame time (INT_TIME + READTIME)
NUM_FRM =                  900 / number of frames per time stamp
NREADOUT=                  720 / number of read per cadence
CHECKSUM= 'cWOAfTM3cTM9cTM9'   / HDU checksum updated 2020-08-05T02:43:40
DATASUM = '645682208'          / data unit checksum updated 2020-08-05T02:43:40
TTYPE1  = 'TIME    '
TFORM1  = 'D       '
TUNIT1  = 'BJD - 2457000, days'
TDISP1  = 'D14.7   '
TTYPE2  = 'TIMECORR'
TFORM2  = 'E       '
TUNIT2  = 'd       '
TDISP2  = 'E13.6   '
TTYPE3  = 'CADENCENO'
TFORM3  = 'J       '
TDISP3  = 'I10     '
TTYPE4  = 'FLUX_RAW'
TFORM4  = 'D       '
TUNIT4  = 'e-/s    '
TDISP4  = 'E26.17  '
TTYPE5  = 'FLUX_RAW_ERR'
TFORM5  = 'D       '
TUNIT5  = 'e-/s    '
TDISP5  = 'E26.17  '
TTYPE6  = 'FLUX_BKG'
TFORM6  = 'D       '
TUNIT6  = 'e-/s    '
TDISP6  = 'E26.17  '
TTYPE7  = 'FLUX_CORR'
TFORM7  = 'D       '
TUNIT7  = 'ppm     '
TDISP7  = 'E26.17  '
TTYPE8  = 'FLUX_CORR_ERR'
TFORM8  = 'D       '
TUNIT8  = 'ppm     '
TDISP8  = 'E26.17  '
TTYPE9  = 'QUALITY '
TFORM9  = 'J       '
TDISP9  = 'B16.16  '
TTYPE10 = 'PIXEL_QUALITY'
TFORM10 = 'J       '
TDISP10 = 'B16.16  '
TTYPE11 = 'MOM_CENTR1'
TFORM11 = 'D       '
TUNIT11 = 'pixels  '
TDISP11 = 'F10.5   '
TTYPE12 = 'MOM_CENTR2'
TFORM12 = 'D       '
TUNIT12 = 'pixels  '
TDISP12 = 'F10.5   '
TTYPE13 = 'POS_CORR1'
TFORM13 = 'D       '
TUNIT13 = 'pixels  '
TDISP13 = 'F14.7   '
TTYPE14 = 'POS_CORR2'
TFORM14 = 'D       '
TUNIT14 = 'pixels  '
TDISP14 = 'F14.7   '
CORRMET = 'CBV     '           / Lightcurve correction method
CORRVER = 'devel-v1.0.15.post221+git9abe1f0' / Version of correction pipeline
CBV_AREA=                  114 / CBV area of star
CBV_MET = 'LS      '           / Method used to fit CBVs
CBV_BIC =                    T / Was BIC used to select no of CBVs
CBV_PRI =                    F / Was prior used
CBV_MAX =                   13 / Number of possible CBVs to fit
CBV_NUM =                   10 / Number of fitted CBVs
CBV_C0  = -0.00015296180968926 / Fitted offset
CBV_C1  =    1.675835455624358 / CBV1 coefficient
CBV_C2  =   0.5158549167363414 / CBV2 coefficient
CBV_C3  =  -0.2279278941917842 / CBV3 coefficient
CBV_C4  =  0.04447041655068518 / CBV4 coefficient
CBV_C5  = 0.009879424483019726 / CBV5 coefficient
CBV_C6  = -0.04301179660182362 / CBV6 coefficient
CBV_C7  =  0.06184856134049641 / CBV7 coefficient
CBV_C8  =  0.01511753191308629 / CBV8 coefficient
CBV_C9  = -0.05409421553225097 / CBV9 coefficient
CBV_C10 = -0.02611387164562197 / CBV10 coefficient
CBVS_C1 =    1.515027165682073 / Spike-CBV1 coefficient
CBVS_C2 =    1.090992296760904 / Spike-CBV2 coefficient
CBVS_C3 =   0.2801410239701041 / Spike-CBV3 coefficient
CBVS_C4 =  -0.1156876403423044 / Spike-CBV4 coefficient
CBVS_C5 =   0.1196523579900063 / Spike-CBV5 coefficient
CBVS_C6 = -0.01563028829168067 / Spike-CBV6 coefficient
CBVS_C7 =  0.04356518394135422 / Spike-CBV7 coefficient
CBVS_C8 = -0.07713142488895069 / Spike-CBV8 coefficient
CBVS_C9 = -0.05557780670250756 / Spike-CBV9 coefficient
CBVS_C10=  0.05747070976331653 / Spike-CBV10 coefficient
END
\end{verbatim}

\subsection*{ENSEMBLE HDU-1 (LIGHTCURVE)}
\begin{verbatim}
XTENSION= 'BINTABLE'           / binary table extension
BITPIX  =                    8 / array data type
NAXIS   =                    2 / number of array dimensions
NAXIS1  =                   96 / length of dimension 1
NAXIS2  =                  993 / length of dimension 2
PCOUNT  =                    0 / number of group parameters
GCOUNT  =                    1 / number of groups
TFIELDS =                   14 / number of table fields
INHERIT =                    T / inherit the primary header
EXTNAME = 'LIGHTCURVE'         / extension name
TIMEREF = 'SOLARSYSTEM'        / barycentric correction applied to times
TIMESYS = 'TDB     '           / time system is Barycentric Dynamical Time (TDB)
BJDREFI =              2457000 / integer part of BTJD reference date
BJDREFF =                  0.0 / fraction of the day in BTJD reference date
TIMEUNIT= 'd       '           / time unit for TIME, TSTART and TSTOP
TSTART  =    1468.276639141453 / observation start time in BTJD
TSTOP   =     1490.04694595576 / observation stop time in BTJD
DATE-OBS= '2018-12-15T18:37:12.438' / TSTART as UTC calendar date
DATE-END= '2019-01-06T13:06:26.946' / TSTOP as UTC calendar date
MJD-BEG =    58467.77663914146 / observation start time in MJD
MJD-END =    58489.54694595576 / observation start time in MJD
TELAPSE =    21.77030681430688 / [d] TSTOP - TSTART
LIVETIME=    4.310520749232762 / [d] TELAPSE multiplied by DEADC
DEADC   =                0.198 / deadtime correction
EXPOSURE=    4.310520749232762 / [d] time on source
XPOSURE =                356.4 / [s] Duration of exposure
TIMEPIXR=                  0.5 / bin time beginning=0 middle=0.5 end=1
TIMEDEL =  0.02083333333333333 / [d] time resolution of data
INT_TIME=                 1.98 / [s] photon accumulation time per frame
READTIME=                 0.02 / [s] readout time per frame
FRAMETIM=                  2.0 / [s] frame time (INT_TIME + READTIME)
NUM_FRM =                  900 / number of frames per time stamp
NREADOUT=                  720 / number of read per cadence
CHECKSUM= 'ZJXgg9XfZGXfd9Xf'   / HDU checksum updated 2020-08-03T19:33:25
DATASUM = '1534575554'         / data unit checksum updated 2020-08-03T19:33:25
TTYPE1  = 'TIME    '
TFORM1  = 'D       '
TUNIT1  = 'BJD - 2457000, days'
TDISP1  = 'D14.7   '
TTYPE2  = 'TIMECORR'
TFORM2  = 'E       '
TUNIT2  = 'd       '
TDISP2  = 'E13.6   '
TTYPE3  = 'CADENCENO'
TFORM3  = 'J       '
TDISP3  = 'I10     '
TTYPE4  = 'FLUX_RAW'
TFORM4  = 'D       '
TUNIT4  = 'e-/s    '
TDISP4  = 'E26.17  '
TTYPE5  = 'FLUX_RAW_ERR'
TFORM5  = 'D       '
TUNIT5  = 'e-/s    '
TDISP5  = 'E26.17  '
TTYPE6  = 'FLUX_BKG'
TFORM6  = 'D       '
TUNIT6  = 'e-/s    '
TDISP6  = 'E26.17  '
TTYPE7  = 'FLUX_CORR'
TFORM7  = 'D       '
TUNIT7  = 'ppm     '
TDISP7  = 'E26.17  '
TTYPE8  = 'FLUX_CORR_ERR'
TFORM8  = 'D       '
TUNIT8  = 'ppm     '
TDISP8  = 'E26.17  '
TTYPE9  = 'QUALITY '
TFORM9  = 'J       '
TDISP9  = 'B16.16  '
TTYPE10 = 'PIXEL_QUALITY'
TFORM10 = 'J       '
TDISP10 = 'B16.16  '
TTYPE11 = 'MOM_CENTR1'
TFORM11 = 'D       '
TUNIT11 = 'pixels  '
TDISP11 = 'F10.5   '
TTYPE12 = 'MOM_CENTR2'
TFORM12 = 'D       '
TUNIT12 = 'pixels  '
TDISP12 = 'F10.5   '
TTYPE13 = 'POS_CORR1'
TFORM13 = 'D       '
TUNIT13 = 'pixels  '
TDISP13 = 'F14.7   '
TTYPE14 = 'POS_CORR2'
TFORM14 = 'D       '
TUNIT14 = 'pixels  '
TDISP14 = 'F14.7   '
CORRMET = 'Ensemble'           / Lightcurve correction method
CORRVER = 'devel-v1.0.15.post221+git9abe1f0' / Version of correction pipeline
ENS_MED =    183.7447509765625 / Median of corrected light curve before ppm
ENS_NUM =                    7 / Number of targets in ensemble
ENS_DLIM=                  1.0 / Limit on differenced range metric
ENS_DREL=                   10 / Limit on relative diff. range
ENS_RLIM=                  0.4 / Limit on flux range metric
END
\end{verbatim}

\subsection*{ENSEMBLE HDU-4 (ENSEMBLE)}
\begin{verbatim}
XTENSION= 'BINTABLE'           / binary table extension
BITPIX  =                    8 / array data type
NAXIS   =                    2 / number of array dimensions
NAXIS1  =                   12 / length of dimension 1
NAXIS2  =                    7 / length of dimension 2
PCOUNT  =                    0 / number of group parameters
GCOUNT  =                    1 / number of groups
TFIELDS =                    2 / number of table fields
TTYPE1  = 'TIC     '           / column title: TIC identifier
TFORM1  = 'K       '           / column format: signed 64-bit integer
TTYPE2  = 'BZETA   '           / column title: background scale
TFORM2  = 'E       '           / column format: 32-bit floating point
EXTNAME = 'ENSEMBLE'           / extension name
TDISP1  = 'I10     '           / column display format
TDISP2  = 'E       '           / column display format
END
\end{verbatim}

\vspace{1cm}
\bibliographystyle{aasjournal}
\bibliography{biblio}

\begin{thebibliography}{}
\expandafter\ifx\csname natexlab\endcsname\relax\def\natexlab#1{#1}\fi
\providecommand{\url}[1]{\href{#1}{#1}}

\bibitem[{{Aigrain} {et~al.}(2008){Aigrain}, {Collier Cameron}, {Ollivier},
  {Pont}, {Jorda}, {Almenara}, {Alonso}, {Barge}, {Bord{\'e}}, {Bouchy},
  {Deeg}, {de La Reza}, {Deleuil}, {Dvorak}, {Erikson}, {Fridlund}, {Gondoin},
  {Gillon}, {Guillot}, {Hatzes}, {Lammer}, {Lanza}, {L{\'e}ger}, {Llebaria},
  {Magain}, {Mazeh}, {Moutou}, {Paetzold}, {Pinte}, {Queloz}, {Rauer}, {Rouan},
  {Schneider}, {Wuchter}, \& {Zucker}}]{corot4}
{Aigrain}, S., {Collier Cameron}, A., {Ollivier}, M., {et~al.} 2008, \aap, 488,
  L43

\bibitem[{{Antoci} {et~al.}(2019){Antoci}, {Cunha}, {Bowman}, {Murphy},
  {Kurtz}, {Bedding}, {Borre}, {Christophe}, {Daszy{\'n}ska-Daszkiewicz},
  {Fox-Machado}, {Garc{\'\i}a Hern{\'a}ndez}, {Ghasemi}, {Handberg}, {Hansen},
  {Hasanzadeh}, {Houdek}, {Johnston}, {Justesen}, {Kahraman Alicavus},
  {Kotysz}, {Latham}, {Matthews}, {M{\o}nster}, {Niemczura}, {Paunzen},
  {S{\'a}nchez Arias}, {Pigulski}, {Pepper}, {Richey-Yowell}, {Safari},
  {Seager}, {Smalley}, {Shutt}, {S{\'o}dor}, {Su{\'a}rez}, {Tkachenko}, {Wu},
  {Zwintz}, {Barcel{\'o} Forteza}, {Brunsden}, {Bogn{\'a}r}, {Buzasi},
  {Chowdhury}, {De Cat}, {Evans}, {Guo}, {Guzik}, {Jevtic}, {Lampens}, {Lares
  Martiz}, {Lovekin}, {Li}, {Mirouh}, {Mkrtichian}, {Monteiro}, {Nemec},
  {Ouazzani}, {Pascual-Granado}, {Reese}, {Rieutord}, {Rodon}, {Skarka},
  {Sowicka}, {Stateva}, {Szab{\'o}}, \& {Weiss}}]{Antoci2019}
{Antoci}, V., {Cunha}, M.~S., {Bowman}, D.~M., {et~al.} 2019, \mnras, 490, 4040

\bibitem[{{Astropy Collaboration} {et~al.}(2013){Astropy Collaboration},
  {Robitaille}, {Tollerud}, {Greenfield}, {Droettboom}, {Bray}, {Aldcroft},
  {Davis}, {Ginsburg}, {Price-Whelan}, {Kerzendorf}, {Conley}, {Crighton},
  {Barbary}, {Muna}, {Ferguson}, {Grollier}, {Parikh}, {Nair}, {Unther},
  {Deil}, {Woillez}, {Conseil}, {Kramer}, {Turner}, {Singer}, {Fox}, {Weaver},
  {Zabalza}, {Edwards}, {Azalee Bostroem}, {Burke}, {Casey}, {Crawford},
  {Dencheva}, {Ely}, {Jenness}, {Labrie}, {Lim}, {Pierfederici}, {Pontzen},
  {Ptak}, {Refsdal}, {Servillat}, \& {Streicher}}]{astropy:2013}
{Astropy Collaboration}, {Robitaille}, T.~P., {Tollerud}, E.~J., {et~al.} 2013,
  \aap, 558, A33

\bibitem[{{Audenaert} {et~al.}(2021){Audenaert}, {Kuszlewicz}, {Handberg},
  {Tkachenko}, {Armstrong}, {Hon}, {Kgoadi}, {Lund}, {Bell}, {Bugnet},
  {Bowman}, {Johnston}, {Garc{\'\i}a}, {Stello}, {Moln{\'a}r}, {Plachy},
  {Buzasi}, {Aerts}, \& {the T'DA collaboration}}]{PaperIII}
{Audenaert}, J., {Kuszlewicz}, J.~S., {Handberg}, R., {et~al.} 2021, arXiv
  e-prints, arXiv:2107.06301

\bibitem[{{Baglin} {et~al.}(2009){Baglin}, {Auvergne}, {Barge}, {Deleuil},
  {Michel}, \& {CoRoT Exoplanet Science Team}}]{corot}
{Baglin}, A., {Auvergne}, M., {Barge}, P., {et~al.} 2009, in IAU Symposium,
  Vol. 253, IAU Symposium, ed. F.~{Pont}, D.~{Sasselov}, \& M.~J. {Holman},
  71--81

\bibitem[{{Basri} {et~al.}(2011){Basri}, {Walkowicz}, {Batalha}, {Gilliland},
  {Jenkins}, {Borucki}, {Koch}, {Caldwell}, {Dupree}, \& {Latham}}]{Basri2011}
{Basri}, G., {Walkowicz}, L.~M., {Batalha}, N., {et~al.} 2011, \aj, 141, 20

\bibitem[{{Bowman}(2017)}]{Bowman2017}
{Bowman}, D.~M. 2017, {Amplitude Modulation of Pulsation Modes in Delta Scuti
  Stars}, doi:10.1007/978-3-319-66649-5

\bibitem[{{Brahm} {et~al.}(2018){Brahm}, {Hartman}, {Jord{\'a}n}, {Bakos},
  {Espinoza}, {Rabus}, {Bhatti}, {Penev}, {Sarkis}, {Suc}, {Csubry}, {Bayliss},
  {Bento}, {Zhou}, {Mancini}, {Henning}, {Ciceri}, {de Val-Borro}, {Shectman},
  {Crane}, {Arriagada}, {Butler}, {Teske}, {Thompson}, {Osip}, {D{\'\i}az},
  {Schmidt}, {L{\'a}z{\'a}r}, {Papp}, \& {S{\'a}ri}}]{HATS45}
{Brahm}, R., {Hartman}, J.~D., {Jord{\'a}n}, A., {et~al.} 2018, \aj, 155, 112

\bibitem[{{Buzasi} {et~al.}(2015){Buzasi}, {Carboneau}, {Hessler}, {Lezcano},
  \& {Preston}}]{Buzasi2015}
{Buzasi}, D.~L., {Carboneau}, L., {Hessler}, C., {Lezcano}, A., \& {Preston},
  H. 2015, in IAU General Assembly, Vol.~29, 2256843

\bibitem[{{Caldwell} {et~al.}(2020){Caldwell}, {Tenenbaum}, {Twicken},
  {Jenkins}, {Ting}, {Smith}, {Hedges}, {Fausnaugh}, {Rose}, \&
  {Burke}}]{2020RNAAS...4..201C}
{Caldwell}, D.~A., {Tenenbaum}, P., {Twicken}, J.~D., {et~al.} 2020, Research
  Notes of the American Astronomical Society, 4, 201

\bibitem[{{Campante} {et~al.}(2016){Campante}, {Schofield}, {Kuszlewicz},
  {Bouma}, {Chaplin}, {Huber}, {Christensen-Dalsgaard}, {Kjeldsen}, {Bossini},
  {North}, {Appourchaux}, {Latham}, {Pepper}, {Ricker}, {Stassun},
  {Vanderspek}, \& {Winn}}]{Campante2016}
{Campante}, T.~L., {Schofield}, M., {Kuszlewicz}, J.~S., {et~al.} 2016, \apj,
  830, 138

\bibitem[{{Chaplin} \& {Miglio}(2013)}]{chaplin2013}
{Chaplin}, W.~J., \& {Miglio}, A. 2013, \araa, 51, 353

\bibitem[{{Feinstein} {et~al.}(2019){Feinstein}, {Montet}, {Foreman-Mackey},
  {Bedell}, {Saunders}, {Bean}, {Christiansen}, {Hedges}, {Luger}, {Scolnic},
  \& {Cardoso}}]{eleanor}
{Feinstein}, A.~D., {Montet}, B.~T., {Foreman-Mackey}, D., {et~al.} 2019,
  \pasp, 131, 094502

\bibitem[{{Garc{\'\i}a} \& {Ballot}(2019)}]{Garcia2019}
{Garc{\'\i}a}, R.~A., \& {Ballot}, J. 2019, Living Reviews in Solar Physics,
  16, 4

\bibitem[{{Garc{\'\i}a} {et~al.}(2014){Garc{\'\i}a}, {Mathur}, {Pires},
  {R{\'e}gulo}, {Bellamy}, {Pall{\'e}}, {Ballot}, {Barcel{\'o} Forteza},
  {Beck}, {Bedding}, {Ceillier}, {Roca Cort{\'e}s}, {Salabert}, \&
  {Stello}}]{rafa2014}
{Garc{\'\i}a}, R.~A., {Mathur}, S., {Pires}, S., {et~al.} 2014, \aap, 568, A10

\bibitem[{{Gilliland} \& {Brown}(1988)}]{brown_gilliland1988}
{Gilliland}, R.~L., \& {Brown}, T.~M. 1988, \pasp, 100, 754

\bibitem[{{Gilliland} {et~al.}(2010){Gilliland}, {Brown},
  {Christensen-Dalsgaard}, {Kjeldsen}, {Aerts}, {Appourchaux}, {Basu},
  {Bedding}, {Chaplin}, {Cunha}, {De Cat}, {De Ridder}, {Guzik}, {Handler},
  {Kawaler}, {Kiss}, {Kolenberg}, {Kurtz}, {Metcalfe}, {Monteiro}, {Szab{\'o}},
  {Arentoft}, {Balona}, {Debosscher}, {Elsworth}, {Quirion}, {Stello},
  {Su{\'a}rez}, {Borucki}, {Jenkins}, {Koch}, {Kondo}, {Latham}, {Rowe}, \&
  {Steffen}}]{Gilliland2010}
{Gilliland}, R.~L., {Brown}, T.~M., {Christensen-Dalsgaard}, J., {et~al.} 2010,
  \pasp, 122, 131

\bibitem[{Handberg {et~al.}(2021{\natexlab{a}})Handberg, Hansen, Pope, \&
  Lund}]{rasmus_handberg_2021_5153073}
Handberg, R., Hansen, J.~S., Pope, B., \& Lund, M.~N. 2021{\natexlab{a}},
  tasoc/photometry: Version 6.2.5, vv6.2.5,  Zenodo,
  doi:10.5281/zenodo.5153073.
\newblock \url{https://doi.org/10.5281/zenodo.5153073}

\bibitem[{Handberg {et~al.}(2021{\natexlab{b}})Handberg, Lund, Carboneau,
  Pereira, \& Reeth}]{rasmus_handberg_2021_5154027}
Handberg, R., Lund, M.~N., Carboneau, L., Pereira, F., \& Reeth, T.~V.
  2021{\natexlab{b}}, tasoc/corrections: Version 2.0.1, vv2.0.1,  Zenodo,
  doi:10.5281/zenodo.5154027.
\newblock \url{https://doi.org/10.5281/zenodo.5154027}

\bibitem[{{Handberg} {et~al.}(2021){Handberg}, {Lund}, {White}, {Hall},
  {Buzasi}, {Pope}, {Hansen}, {von Essen}, {Carboneau}, {Huber}, {Vanderspek},
  {Fausnaug}, {Tenenbaum}, {Jenkins}, \& {the T'DA Collaboration}}]{PaperI}
{Handberg}, R., {Lund}, M.~N., {White}, T.~R., {et~al.} 2021, arXiv e-prints,
  arXiv:2106.08341

\bibitem[{Harris {et~al.}(2020)Harris, Millman, van~der Walt, Gommers,
  Virtanen, Cournapeau, Wieser, Taylor, Berg, Smith, Kern, Picus, Hoyer, van
  Kerkwijk, Brett, Haldane, del R{'{\i}}o, Wiebe, Peterson,
  G{'{e}}rard-Marchant, Sheppard, Reddy, Weckesser, Abbasi, Gohlke, \&
  Oliphant}]{harris2020array}
Harris, C.~R., Millman, K.~J., van~der Walt, S.~J., {et~al.} 2020, Nature, 585,
  357.
\newblock \url{https://doi.org/10.1038/s41586-020-2649-2}

\bibitem[{{Hartman} {et~al.}(2015){Hartman}, {Bayliss}, {Brahm}, {Bakos},
  {Mancini}, {Jord{\'a}n}, {Penev}, {Rabus}, {Zhou}, {Butler}, {Espinoza}, {de
  Val-Borro}, {Bhatti}, {Csubry}, {Ciceri}, {Henning}, {Schmidt}, {Arriagada},
  {Shectman}, {Crane}, {Thompson}, {Suc}, {Cs{\'a}k}, {Tan}, {Noyes},
  {L{\'a}z{\'a}r}, {Papp}, \& {S{\'a}ri}}]{HATS6}
{Hartman}, J.~D., {Bayliss}, D., {Brahm}, R., {et~al.} 2015, \aj, 149, 166

\bibitem[{{H{\'e}brard} {et~al.}(2011){H{\'e}brard}, {Evans}, {Alonso},
  {Fridlund}, {Ofir}, {Aigrain}, {Guillot}, {Almenara}, {Auvergne}, {Baglin},
  {Barge}, {Bonomo}, {Bord{\'e}}, {Bouchy}, {Cabrera}, {Carone}, {Carpano},
  {Cavarroc}, {Csizmadia}, {Deeg}, {Deleuil}, {D{\'\i}az}, {Dvorak}, {Erikson},
  {Ferraz-Mello}, {Gand olfi}, {Gibson}, {Gillon}, {Guenther}, {Hatzes},
  {Havel}, {Jorda}, {Lammer}, {L{\'e}ger}, {Llebaria}, {Mazeh}, {Moutou},
  {Ollivier}, {Parviainen}, {P{\"a}tzold}, {Queloz}, {Rauer}, {Rouan},
  {Santerne}, {Schneider}, {Tingley}, \& {Wuchterl}}]{corot18-2011}
{H{\'e}brard}, G., {Evans}, T.~M., {Alonso}, R., {et~al.} 2011, \aap, 533, A130

\bibitem[{Hekker \& Christensen-Dalsgaard(2017)}]{Hekker2017}
Hekker, S., \& Christensen-Dalsgaard, J. 2017, The Astronomy and Astrophysics
  Review, 25, 1.
\newblock \url{https://doi.org/10.1007/s00159-017-0101-x}

\bibitem[{{Holdsworth} {et~al.}(2021){Holdsworth}, {Cunha}, {Kurtz}, {Antoci},
  {Hey}, {Bowman}, {Kobzar}, {Buzasi}, {Kochukhov}, {Niemczura}, {Ozuyar},
  {Shi}, {Szab{\'o}}, {Samadi-Ghadim}, {Bogn{\'a}r}, {Fox-Machado}, {Khalack},
  {Lares-Martiz}, {Lovekin}, {Miko{\l}ajczyk}, {Mkrtichian}, {Pascual-Granado},
  {Paunzen}, {Richey-Yowell}, {S{\'o}dor}, {Sikora}, {Yang}, {Brunsden},
  {David-Uraz}, {Derekas}, {Garc{\'\i}a Hern{\'a}ndez}, {Guzik}, {Hatamkhani},
  {Handberg}, {Lambert}, {Lampens}, {Murphy}, {Monier}, {Pollard},
  {Quitral-Manosalva}, {Ram{\'o}n-Ballesta}, {Smalley}, {Stateva}, \&
  {Vanderspek}}]{Holdsworth2021}
{Holdsworth}, D.~L., {Cunha}, M.~S., {Kurtz}, D.~W., {et~al.} 2021, \mnras,
  506, 1073

\bibitem[{{Howell} {et~al.}(2014){Howell}, {Sobeck}, {Haas}, {Still},
  {Barclay}, {Mullally}, {Troeltzsch}, {Aigrain}, {Bryson}, {Caldwell},
  {Chaplin}, {Cochran}, {Huber}, {Marcy}, {Miglio}, {Najita}, {Smith},
  {Twicken}, \& {Fortney}}]{Howell2014}
{Howell}, S.~B., {Sobeck}, C., {Haas}, M., {et~al.} 2014, \pasp, 126, 398

\bibitem[{{Huang} {et~al.}(2020){Huang}, {Vanderburg}, {P{\'a}l}, {Sha}, {Yu},
  {Fong}, {Fausnaugh}, {Shporer}, {Guerrero}, {Vanderspek}, \& {Ricker}}]{qlp2}
{Huang}, C.~X., {Vanderburg}, A., {P{\'a}l}, A., {et~al.} 2020, Research Notes
  of the American Astronomical Society, 4, 206

\bibitem[{{Huber} {et~al.}(2011){Huber}, {Bedding}, {Stello}, {Hekker},
  {Mathur}, {Mosser}, {Verner}, {Bonanno}, {Buzasi}, {Campante}, {Elsworth},
  {Hale}, {Kallinger}, {Silva Aguirre}, {Chaplin}, {De Ridder},
  {Garc{\'{\i}}a}, {Appourchaux}, {Frandsen}, {Houdek}, {Molenda-{\.Z}akowicz},
  {Monteiro}, {Christensen-Dalsgaard}, {Gilliland}, {Kawaler}, {Kjeldsen},
  {Broomhall}, {Corsaro}, {Salabert}, {Sanderfer}, {Seader}, \&
  {Smith}}]{huber2011}
{Huber}, D., {Bedding}, T.~R., {Stello}, D., {et~al.} 2011, \apj, 743, 143

\bibitem[{Hunter(2011)}]{matplot}
Hunter, J.~D. 2011, Computing in Science \& Engineering, 9, 90

\bibitem[{{Jenkins} {et~al.}(2016){Jenkins}, {Twicken}, {McCauliff},
  {Campbell}, {Sanderfer}, {Lung}, {Mansouri-Samani}, {Girouard}, {Tenenbaum},
  {Klaus}, {Smith}, {Caldwell}, {Chacon}, {Henze}, {Heiges}, {Latham},
  {Morgan}, {Swade}, {Rinehart}, \& {Vanderspek}}]{SPOC}
{Jenkins}, J.~M., {Twicken}, J.~D., {McCauliff}, S., {et~al.} 2016, in Society
  of Photo-Optical Instrumentation Engineers (SPIE) Conference Series, Vol.
  9913, Software and Cyberinfrastructure for Astronomy IV, ed. G.~{Chiozzi} \&
  J.~C. {Guzman}, 99133E

\bibitem[{Jones {et~al.}(2001)Jones, Oliphant, Peterson, {et~al.}}]{Scipy}
Jones, E., Oliphant, T., Peterson, P., {et~al.} 2001, {SciPy}: Open source
  scientific tools for {Python}, , .
\newblock \url{http://www.scipy.org/}

\bibitem[{{Jord{\'a}n} {et~al.}(2014){Jord{\'a}n}, {Brahm}, {Bakos}, {Bayliss},
  {Penev}, {Hartman}, {Zhou}, {Mancini}, {Mohler-Fischer}, {Ciceri}, {Sato},
  {Csubry}, {Rabus}, {Suc}, {Espinoza}, {Bhatti}, {de Val-Borro}, {Buchhave},
  {Cs{\'a}k}, {Henning}, {Schmidt}, {Tan}, {Noyes}, {B{\'e}ky}, {Butler},
  {Shectman}, {Crane}, {Thompson}, {Williams}, {Martin}, {Contreras},
  {L{\'a}z{\'a}r}, {Papp}, \& {S{\'a}ri}}]{HATS4}
{Jord{\'a}n}, A., {Brahm}, R., {Bakos}, G.~{\'A}., {et~al.} 2014, \aj, 148, 29

\bibitem[{{Lightkurve Collaboration} {et~al.}(2018{\natexlab{a}}){Lightkurve
  Collaboration}, {Cardoso}, {Hedges}, {Gully-Santiago}, {Saunders}, {Cody},
  {Barclay}, {Hall}, {Sagear}, {Turtelboom}, {Zhang}, {Tzanidakis}, {Mighell},
  {Coughlin}, {Bell}, {Berta-Thompson}, {Williams}, {Dotson}, \&
  {Barentsen}}]{lightkurve}
{Lightkurve Collaboration}, {Cardoso}, J.~V.~d.~M., {Hedges}, C., {et~al.}
  2018{\natexlab{a}}, {Lightkurve: Kepler and TESS time series analysis in
  Python}, Astrophysics Source Code Library, , , ascl:1812.013

\bibitem[{{Lightkurve Collaboration} {et~al.}(2018{\natexlab{b}}){Lightkurve
  Collaboration}, {Cardoso}, {Hedges}, {Gully-Santiago}, {Saunders}, {Cody},
  {Barclay}, {Hall}, {Sagear}, {Turtelboom}, {Zhang}, {Tzanidakis}, {Mighell},
  {Coughlin}, {Bell}, {Berta-Thompson}, {Williams}, {Dotson}, \&
  {Barentsen}}]{2018ascl.soft12013L}
---. 2018{\natexlab{b}}, {Lightkurve: Kepler and TESS time series analysis in
  Python}, Astrophysics Source Code Library, , , ascl:1812.013

\bibitem[{{Lund} {et~al.}(2017){Lund}, {Handberg}, {Kjeldsen}, {Chaplin}, \&
  {Christensen-Dalsgaard}}]{Lund2017}
{Lund}, M.~N., {Handberg}, R., {Kjeldsen}, H., {Chaplin}, W.~J., \&
  {Christensen-Dalsgaard}, J. 2017, in European Physical Journal Web of
  Conferences, Vol. 160, European Physical Journal Web of Conferences, 01005

\bibitem[{{Miglio} {et~al.}(2013){Miglio}, {Chiappini}, {Morel}, {Barbieri},
  {Chaplin}, {Girardi}, {Montalb{\'a}n}, {Valentini}, {Mosser}, {Baudin},
  {Casagrande}, {Fossati}, {Silva Aguirre}, \& {Baglin}}]{Miglio2013}
{Miglio}, A., {Chiappini}, C., {Morel}, T., {et~al.} 2013, \mnras, 429, 423

\bibitem[{Nason(2006)}]{Nason2006}
Nason, G. 2006, IAVCEI Spec. Pub. 1, 129

\bibitem[{{Oelkers} \& {Stassun}(2018)}]{Oelkers2018}
{Oelkers}, R.~J., \& {Stassun}, K.~G. 2018, \aj, 156, 132

\bibitem[{{Pascual-Granado} {et~al.}(2018){Pascual-Granado}, {Su{\'a}rez},
  {Garrido}, {Moya}, {Garc{\'\i}a Hern{\'a}ndez}, {Rod{\'o}n}, \&
  {Lares-Martiz}}]{2018A&A...614A..40P}
{Pascual-Granado}, J., {Su{\'a}rez}, J.~C., {Garrido}, R., {et~al.} 2018, \aap,
  614, A40

\bibitem[{Pedregosa {et~al.}(2011)Pedregosa, Varoquaux, Gramfort, Michel,
  Thirion, Grisel, Blondel, Prettenhofer, Weiss, Dubourg, Vanderplas, Passos,
  Cournapeau, Brucher, Perrot, \& Duchesnay}]{scikit-learn}
Pedregosa, F., Varoquaux, G., Gramfort, A., {et~al.} 2011, Journal of Machine
  Learning Research, 12, 2825

\bibitem[{{Price-Whelan} {et~al.}(2018){Price-Whelan}, {Sip{\H{o}}cz},
  {G{\"u}nther}, {Lim}, {Crawford}, {Conseil}, {Shupe}, {Craig}, {Dencheva},
  {Ginsburg}, {VanderPlas}, {Bradley}, {P{\'e}rez-Su{\'a}rez}, {de Val-Borro},
  {Paper Contributors}, {Aldcroft}, {Cruz}, {Robitaille}, {Tollerud},
  {Coordination Committee}, {Ardelean}, {Babej}, {Bach}, {Bachetti}, {Bakanov},
  {Bamford}, {Barentsen}, {Barmby}, {Baumbach}, {Berry}, {Biscani}, {Boquien},
  {Bostroem}, {Bouma}, {Brammer}, {Bray}, {Breytenbach}, {Buddelmeijer},
  {Burke}, {Calderone}, {Cano Rodr{\'\i}guez}, {Cara}, {Cardoso}, {Cheedella},
  {Copin}, {Corrales}, {Crichton}, {D{\textquoteright}Avella}, {Deil},
  {Depagne}, {Dietrich}, {Donath}, {Droettboom}, {Earl}, {Erben}, {Fabbro},
  {Ferreira}, {Finethy}, {Fox}, {Garrison}, {Gibbons}, {Goldstein}, {Gommers},
  {Greco}, {Greenfield}, {Groener}, {Grollier}, {Hagen}, {Hirst}, {Homeier},
  {Horton}, {Hosseinzadeh}, {Hu}, {Hunkeler}, {Ivezi{\'c}}, {Jain}, {Jenness},
  {Kanarek}, {Kendrew}, {Kern}, {Kerzendorf}, {Khvalko}, {King}, {Kirkby},
  {Kulkarni}, {Kumar}, {Lee}, {Lenz}, {Littlefair}, {Ma}, {Macleod},
  {Mastropietro}, {McCully}, {Montagnac}, {Morris}, {Mueller}, {Mumford},
  {Muna}, {Murphy}, {Nelson}, {Nguyen}, {Ninan}, {N{\"o}the}, {Ogaz}, {Oh},
  {Parejko}, {Parley}, {Pascual}, {Patil}, {Patil}, {Plunkett}, {Prochaska},
  {Rastogi}, {Reddy Janga}, {Sabater}, {Sakurikar}, {Seifert}, {Sherbert},
  {Sherwood-Taylor}, {Shih}, {Sick}, {Silbiger}, {Singanamalla}, {Singer},
  {Sladen}, {Sooley}, {Sornarajah}, {Streicher}, {Teuben}, {Thomas},
  {Tremblay}, {Turner}, {Terr{\'o}n}, {van Kerkwijk}, {de la Vega}, {Watkins},
  {Weaver}, {Whitmore}, {Woillez}, {Zabalza}, \& {Contributors}}]{astropy:2018}
{Price-Whelan}, A.~M., {Sip{\H{o}}cz}, B.~M., {G{\"u}nther}, H.~M., {et~al.}
  2018, \aj, 156, 123

\bibitem[{{Pr{\v{s}}a} {et~al.}(2019){Pr{\v{s}}a}, {Zhang}, \&
  {Wells}}]{Prsa2019}
{Pr{\v{s}}a}, A., {Zhang}, M., \& {Wells}, M. 2019, \pasp, 131, 068001

\bibitem[{{Raetz} {et~al.}(2019){Raetz}, {Heras}, {Fern{\'a}ndez}, {Casanova},
  \& {Marka}}]{Raetz2019}
{Raetz}, S., {Heras}, A.~M., {Fern{\'a}ndez}, M., {Casanova}, V., \& {Marka},
  C. 2019, \mnras, 483, 824

\bibitem[{{Ricker} {et~al.}(2014){Ricker}, {Winn}, {Vanderspek}, {Latham},
  {Bakos}, {Bean}, {Berta-Thompson}, {Brown}, {Buchhave}, {Butler}, {Butler},
  {Chaplin}, {Charbonneau}, {Christensen-Dalsgaard}, {Clampin}, {Deming},
  {Doty}, {De Lee}, {Dressing}, {Dunham}, {Endl}, {Fressin}, {Ge}, {Henning},
  {Holman}, {Howard}, {Ida}, {Jenkins}, {Jernigan}, {Johnson}, {Kaltenegger},
  {Kawai}, {Kjeldsen}, {Laughlin}, {Levine}, {Lin}, {Lissauer}, {MacQueen},
  {Marcy}, {McCullough}, {Morton}, {Narita}, {Paegert}, {Palle}, {Pepe},
  {Pepper}, {Quirrenbach}, {Rinehart}, {Sasselov}, {Sato}, {Seager},
  {Sozzetti}, {Stassun}, {Sullivan}, {Szentgyorgyi}, {Torres}, {Udry}, \&
  {Villasenor}}]{Ricker2014}
{Ricker}, G.~R., {Winn}, J.~N., {Vanderspek}, R., {et~al.} 2014, in Society of
  Photo-Optical Instrumentation Engineers (SPIE) Conference Series, Vol. 9143,
  Society of Photo-Optical Instrumentation Engineers (SPIE) Conference Series,
  20

\bibitem[{Schofield {et~al.}(2019)Schofield, Chaplin, Huber, Campante, Davies,
  Miglio, Ball, Appourchaux, Basu, Bedding, Christensen-Dalsgaard, Creevey,
  Garc{\'{\i}}a, Handberg, Kawaler, Kjeldsen, Latham, Lund, Metcalfe, Ricker,
  Serenelli, Aguirre, Stello, \& Vanderspek}]{Schofield2019}
Schofield, M., Chaplin, W.~J., Huber, D., {et~al.} 2019, The Astrophysical
  Journal Supplement Series, 241, 12.
\newblock \url{https://doi.org/10.3847%2F1538-4365%2Fab04f5}

\bibitem[{{Silva Aguirre} {et~al.}(2020){Silva Aguirre}, {Stello}, {Stokholm},
  {Mosumgaard}, {Ball}, {Basu}, {Bossini}, {Bugnet}, {Buzasi}, {Campante},
  {Carboneau}, {Chaplin}, {Corsaro}, {Davies}, {Elsworth}, {Garc{\'\i}a},
  {Gaulme}, {Hall}, {Handberg}, {Hon}, {Kallinger}, {Kang}, {Lund}, {Mathur},
  {Mints}, {Mosser}, {{\c{C}}elik Orhan}, {Rodrigues}, {Vrard}, {Y{\i}ld{\i}z},
  {Zinn}, {{\"O}rtel}, {Beck}, {Bell}, {Guo}, {Jiang}, {Kuszlewicz}, {Kuehn},
  {Li}, {Lundkvist}, {Pinsonneault}, {Tayar}, {Cunha}, {Hekker}, {Huber},
  {Miglio}, {F.~G. Monteiro}, {Slumstrup}, {Winther}, {Angelou}, {Benomar},
  {B{\'o}di}, {De Moura}, {Deheuvels}, {Derekas}, {Di Mauro}, {Dupret},
  {Jim{\'e}nez}, {Lebreton}, {Matthews}, {Nardetto}, {do Nascimento},
  {Pereira}, {Rodr{\'\i}guez D{\'\i}az}, {Serenelli}, {Spitoni},
  {Stonkut{\.{e}}}, {Su{\'a}rez}, {Szab{\'o}}, {Van Eylen}, {Ventura}, {Verma},
  {Weiss}, {Wu}, {Barclay}, {Christensen-Dalsgaard}, {Jenkins}, {Kjeldsen},
  {Ricker}, {Seager}, \& {Vanderspek}}]{TESSgiant}
{Silva Aguirre}, V., {Stello}, D., {Stokholm}, A., {et~al.} 2020, \apjl, 889,
  L34

\bibitem[{Smith {et~al.}(2017)Smith, Stumpe, M., C., R., J., D., L., \&
  D.}]{kepdatc8}
Smith, J.~C., Stumpe, M.~C., M., J.~J., {et~al.} 2017, “Presearch data
  conditioning,” in Kepler Data Processing Handbook: KSCI-19081-002, Jenkins,
  J. M. (ed.), ,

\bibitem[{{Stumpe} {et~al.}(2012){Stumpe}, {Smith}, {Van Cleve}, {Twicken},
  {Barclay}, {Fanelli}, {Girouard}, {Jenkins}, {Kolodziejczak}, {McCauliff}, \&
  {Morris}}]{pdc1}
{Stumpe}, M.~C., {Smith}, J.~C., {Van Cleve}, J.~E., {et~al.} 2012, \pasp, 124,
  985

\bibitem[{{Sullivan} {et~al.}(2015){Sullivan}, {Winn}, {Berta-Thompson},
  {Charbonneau}, {Deming}, {Dressing}, {Latham}, {Levine}, {McCullough}, \&
  {Morton}}]{Sullivan2015}
{Sullivan}, P.~W., {Winn}, J.~N., {Berta-Thompson}, Z.~K., {et~al.} 2015, \apj,
  809, 77

\bibitem[{Waskom \& the seaborn~development team(2020)}]{waskom2020seaborn}
Waskom, M., \& the seaborn~development team. 2020, mwaskom/seaborn, vlatest,
  Zenodo, doi:10.5281/zenodo.592845.
\newblock \url{https://doi.org/10.5281/zenodo.592845}

\bibitem[{{White} {et~al.}(2017){White}, {Pope}, {Antoci}, {P{\'a}pics},
  {Aerts}, {Gies}, {Gordon}, {Huber}, {Schaefer}, {Aigrain}, {Albrecht},
  {Barclay}, {Barentsen}, {Beck}, {Bedding}, {Fredslund Andersen}, {Grundahl},
  {Howell}, {Ireland}, {Murphy}, {Nielsen}, {Silva Aguirre}, \&
  {Tuthill}}]{White2017}
{White}, T.~R., {Pope}, B.~J.~S., {Antoci}, V., {et~al.} 2017, \mnras, 471,
  2882

\end{thebibliography}

\end{document}